# Nonlinear susceptibility for stimulated light scattering in plasmas

Aleksei Zheltikov

Institute for Quantum Science and Engineering, Department of Physics and Astronomy, Texas A&M University, College Station, Texas 77843; zheltikov@tamu.edu

A self-consistent treatment of strong-field laser – plasma interactions suggests a heuristically valuable extension of the notion of nonlinear-optical susceptibility to perturbative regimes of stimulated light scattering in plasmas, offering powerful insights into stimulated-scattering-induced cross-beam energy transfer processes. We derive a closed-form, physically intuitive solution for the third-order susceptibility $\chi_p^{(3)}$ for stimulated Brillouin and Raman scattering (SBS and SRS) in plasmas, expressing $\chi_p^{(3)}$ in terms of individual susceptibilities $\chi_e$ and $\chi_i$ of electron and ion plasma constituents. While the individual susceptibilities $\chi_e$ and $\chi_i$ are functions of plasma parameters, depending on the specific regime of laser – plasma interactions, the algebra whereby $\chi_e$ and $\chi_i$ are mixed into $\chi_p^{(3)}$ remains unchanged in a vast variety of physical settings, revealing significant properties of plasma SBS and SRS, including a physically substantiative connection between SBS/SRS spectral lineshapes and Landau damping, and allowing the ideal-fluid cubic susceptibility $\chi_p^{(3)}$ to be understood as a low-order rational approximation to the kinetic-theory solution for $\chi_p^{(3)}$ derived via a specific closure of the hierarchy of equations for the moments of the governing kinetic equation. We show that, in its weak-field, steady-state limit, the general physical picture of plasma SBS and SRS that views these processes as Stokes-field – plasma-wave decay instabilities of the pump field, reduces to a perturbative picture, familiar from nonlinear optics of neutral media, in which constitutive relations for plasma SBS/SRS response are formulated in terms of the third-order nonlinear polarization and the SBS/SRS gain is defined by the imaginary part of $\chi_p^{(3)}$. These findings lay grounds for a unified treatment of stimulated light scattering across a vast area of optical physics, spanning from fiber optics and neutral gases to laser – plasma physics.

## 1. Introduction

Stimulated scattering and parametric transformation of electromagnetic radiation in plasmas are among the most salient manifestations of nonlinear plasma electrodynamics. Plasma nonlinearities behind such processes had been the focus of intense research well before the advent of high-power lasers [1 – 3], motivated largely by the progress in microwave technologies, ideas of microwave-driven plasma physics, as well as problems related to the propagation of electromagnetic waves in ionosphere and space plasmas [4, 5]. These early studies have been pivotal in recognizing the significance of plasma-wave-enhanced parametric instabilities for the physics of interactions of high-intensity radiation with plasmas [6 – 17].

In the era of ultrafast laser science and rapidly progressing high-power laser technologies, stimulated light scattering processes in plasmas, such as stimulated Raman and Brillouin scattering (SRS and SBS), suggest means for the amplification and compression of laser pulses at the levels of field intensities well above the laser-damage threshold of any nonplasma material (see, e.g., Refs. 18 – 51 and Refs. 19, 52 – 74 for, respectively, SRS- and SBS-based approaches). In strong-field laser – matter interactions physics, on the other hand, SBS is often a detrimental side effect that prevents the desired field intensities from being achieved on target surfaces [14 – 17, 75, 76], especially in multibeam laser – target interaction arrangements. In such settings, SBS has been found to give rise to cross-beam energy transfer (CBET) [77 – 94], thus limiting the intensity of laser radiation that can be achieved on a target without breaking the symmetry of beam arrangement. Specifically, SBS-induced CBET has been identified as one of the major roadblocks on the way toward laser-driven inertial confinement fusion [75, 80, 81, 83, 86, 88, 93, 94].

In strong-field laser – plasma physics, stimulated light scattering is described in the framework of linear stability analysis and is understood [1 – 17] as a parametric decay instability of a laser field with respect to the buildup of the Stokes-shifted electromagnetic field and a plasma wave – a low-frequency ion-acoustic wave (IAW) in SBS or an electron Langmuir wave in SRS. Such a treatment has been proven adequate over a vast parameter space, enabling an accurate description of a remarkably a broad variety of laser – plasma stimulated scattering settings and scenarios, including ultrafast stimulated scattering [56 – 63, 67, 70], as well as strong-field SBS and SRS [15 – 17, 58 – 63], in which intense laser fields can significantly alter the properties of collective plasma waves, shifting their frequencies relative to the frequencies of the respective IAW and Langmuir-wave plasma eigenmodes.

In moderate-intensity regimes, on the other hand, the properties of plasma SBS and SRS are adequately understood in terms of interaction of the laser field with IAW and Langmuir-wave eigenmodes [15 – 17, 75, 88, 95 – 98], whose frequencies define the peaks in SBS and SRS gain spectra. While the typical level of laser intensities in such experiments is significantly lower than the level of laser intensities at which the laser field becomes strongly coupled to plasma waves, altering their properties and shifting their frequencies, these intensities are still very high when viewed on the scale of typical laser intensities in atomic and molecular optics, extending well into the realm of laser-driven inertial fusion [93 – 98].

One of the central goals of this study is to explore whether the description of such moderate-intensity regimes of stimulated light scattering in plasmas can be uniformly grounded in the constituent relations operating with nonlinear optical susceptibilities, thus allowing nonlinear-susceptibility-based approaches to be extended to laser – plasma physics. The concept of nonlinear susceptibilities is central to nonlinear optics [99 – 102], serving to express the nonlinear polarization as a perturbative series in optical fields. In laser – plasma studies, methods based on nonlinear susceptibilities are widely used and are highly productive [103 – 109] but are largely limited to the interface between plasma physics and the physics of highly excited, partially ionized gases, including the laser diagnostics of flames, combustion, and low-temperature gas-discharge and laser-induced plasmas [110 – 114]. Nonlinear-susceptibility-based models and methods have been proven central to the development of laser techniques for high-spectral-resolution remote sensing and high-spatial-resolution three-dimensional imaging of ionized gases and plasmas using four-wave mixing, coherent Raman scattering, as well as second- and third-harmonic generation [115 – 119].

In a broader perspective, however, extension of susceptibility-based approaches to nonlinear processes in plasmas meets fundamental difficulties. Unlike atomic and molecular gases, liquids, and solid dielectrics, where material polarization originates from bound-electron orbitals distorted by the driver field, giving rise to local field-induced electric dipoles, the plasma response to an electromagnetic driver is inherently collective and is, therefore, fundamentally nonlocal. The properties of charged particles in plasmas are very different from the properties of bound electrons in atomic and molecular systems. Plasma particles can routinely travel over macroscopic distances, generating new, self-consistent fields and giving rise to plasma waves, as a part of their collective response to an external field [120 – 122]. The nonlinear-optical response of plasmas is thus a function of collective plasma properties rather than properties that could be related to local field-induced dipoles [100, 102, 123]. Analysis of nonlinear optical processes in plasmas should therefore go way beyond the optics of local bound-state dipoles en route to an adequate description of the field-driven collective plasma response. To enable such a treatment, the field-evolution equations for the laser driver need to be solved jointly with the kinetic equations for the probability distribution functions of plasma particles [2 – 5, 15, 85, 95].

Even in the realm of perturbative nonlinear plasma optics, laser-driven nonlinear processes in plasmas are often more adequately understood in terms of nonlinear currents rather than nonlinear polarization [124 – 127]. While the analysis of nonlinear susceptibilities runs into difficulties related to the description of bound – free and free – free transitions, nonlinear plasma currents are often rightfully viewed as an adequate representation for the sources of plasma nonlinearities, meaningfully connecting to the properties of nonlinear-optical signals in laser – plasma studies [100, 102, 128, 129]. Moreover, in ultrafast laser – plasma interactions, the nonlinear plasma response is highly nonstationary, necessitating adequate dynamic description in terms of time-resolving models, often rendering frequency-domain susceptibilities irrelevant. Finally, in very strong laser fields, laser – plasma nonlinear optics is intrinsically nonperturbative [130 – 132], as clearly indicated by the spectra of high-order harmonics and the properties of strong-field stimulated light scattering, thus ruling out any idea of perturbative treatment and related nonlinear-susceptibility-based description.

Here, we show, however, that, within a vast parameter space of laser – plasma physics, including some of the practically significant regimes of strong-field laser – plasma interactions, the notion of nonlinear-optical susceptibility permits a heuristically valuable extension to stimulated light scattering in plasmas, offering powerful insights into the properties of stimulated Brillouin and stimulated Raman scattering (SBS and SRS) from plasma waves and shedding new light on the inner workings of the related cross-beam energy transfer (CBET) processes. We derive a closed-form, physically intuitive solution for the third-order susceptibility $\chi_p^{(3)}$ for stimulated Brillouin and stimulated Raman scattering, expressing $\chi_p^{(3)}$ in terms of individual susceptibilities $\chi_e$ and $\chi_i$ of electron and ion plasma constituents. $\chi_p^{(3)}$ analysis of laser – plasma interactions presented below reveals a physically substantiative connection between SBS/SRS spectral lineshapes and Landau damping and enables a unified treatment and a physically meaningful benchmarking of plasma SBS and SRS against stimulated scattering in atomic, molecular, condensed-matter, and waveguide optics.

**2. The physical framework: evolution of the laser field and plasma response**

We consider a standard setting of stimulated light scattering in laser – plasma interactions [100, 102], in which a laser driver

$$\mathbf{A}(\mathbf{r},t)=\frac{1}{2}\mathbf{A}_0\exp[i(\mathbf{k}_0\cdot\mathbf{r}-\omega_0 t)]+\frac{1}{2}\mathbf{A}_1\exp[i(\mathbf{k}_1\cdot\mathbf{r}-\omega_1 t)]+\text{c.c.}\quad, (1)$$

consisting of pump and Stokes fields with frequencies $\omega_0$ and $\omega_1$, wave vectors $\mathbf{k}_0$ and $\mathbf{k}_1$, and transverse vector-potential amplitudes $\mathbf{A}_0$ and $\mathbf{A}_1$, drives a beat wave with a frequency $\Omega = \omega_0-\omega_1$, wave vector $\mathbf{q}=\mathbf{k}_0-\mathbf{k}_1$, and amplitude $\mathbf{E}_0\mathbf{E}_1^*\exp[i(\mathbf{q}\cdot\mathbf{r}-\Omega t)]$, $\mathbf{E}_{0.1} = (i\omega_{0,1}/c)\mathbf{A}_{0,1}$, exerting a ponderomotive force on plasma particles of sort $\alpha$ with mass $m_\alpha$ and electric charge $q_\alpha$,

$$\mathbf{F}_\alpha(\mathbf{r},t)=-\nabla\widetilde{U}_\alpha(\mathbf{r},t)\ , \quad (2)$$

with

$$\widetilde{U}_\alpha(\mathbf{r},t)=\frac{1}{2}U_\alpha(\mathbf{q},\Omega)\exp[i(\mathbf{q}\cdot\mathbf{r}-\Omega t)]+\text{c.c.}\quad, \quad (3)$$

$$U_\alpha(\mathbf{q},\Omega)=\frac{q_\alpha^2}{2m_\alpha\omega_0\omega_1}\mathbf{E}_0\cdot\mathbf{E}_1^*\quad. \quad (4)$$

When the laser-driven beat wave becomes resonant to one of the plasma-wave eigenmodes, such a laser – plasma interaction opens a channel for efficient energy transfer between the pump and Stokes laser beams, as well as between the laser field and the plasma wave. As one prominent manifestation of such energy transfer, the laser pump combines with resonantly driven Langmuir or ion-acoustic plasma waves to give rise to a rapid growth of the Stokes field via, respectively, stimulated Raman or stimulated Brillouin scattering.

Analysis of such processes involves solving suitable field-evolution equations jointly with equations describing the material response to an electromagnetic driver field subject to continuity equation and pertinent conservation laws [13 – 17, 75, 76]. In a vast variety of practically significant laser – plasma interaction settings in nonrelativistic, unmagnetized plasmas, stimulated light scattering is adequately described [15 – 17, 75, 76] in terms of the wave equation for the transverse vector potential $\mathbf{A}$,

$$\left(\partial^2/\partial t^2 - c^2\nabla^2 + \omega_{pe}^2\right)\mathbf{a} = -\omega_{pe}^2(\delta\tilde{n}_e/n_{e0})\mathbf{a}, \tag{5}$$

where $\mathbf{a} = [e/(m_e c^2)]\mathbf{A}$, $\omega_{pe}^2 = 4\pi e^2 n_{e0}/m_e$ is the electron plasma frequency, $m_e$ and $e$ are the electron mass and charge, $n_{e0}$ is the background, unperturbed electron density, and $\delta\tilde{n}_e$ is the change in the electron density induced by the laser field, such that the total electron density is $n_e = n_{e0} + \delta\tilde{n}_e$ with $|\delta\tilde{n}_e| \ll n_{e0}$.

The framework for plasma-response description [13 – 17, 75, 76] includes a fluid-dynamics equation for the velocity $\tilde{\mathbf{u}}_\alpha(\mathbf{r}, t)$ of plasma particles $\alpha$,

$$\partial\tilde{\mathbf{u}}_\alpha/\partial t + q_\alpha\nabla\tilde{\phi}/m_\alpha + \gamma_\alpha v_{T\alpha}^2\,\nabla(\delta\tilde{n}_\alpha)/n_{\alpha 0} = 0, \tag{6}$$

and continuity equation

$$\partial(\delta\tilde{n}_\alpha)/\partial t + n_{\alpha 0}\nabla\cdot(\tilde{\mathbf{u}}_\alpha) = 0. \tag{7}$$

Here, $n_{\alpha 0}$ is the background density of plasma particles of sort $\alpha$ with mass $m_\alpha$ and electric charge $q_\alpha$, $\delta\tilde{n}_\alpha = n_\alpha - n_{\alpha 0}$ is the change in the density $n_\alpha$ of plasma particles $\alpha$ relative to its background level $n_{\alpha 0} \gg |\delta\tilde{n}_\alpha|$ due to the external force, $v_{T\alpha} = \sqrt{k_B T_\alpha/m_\alpha}$ is the thermal velocity of plasma particles $\alpha$ at temperature $T_\alpha$, $k_B$ is the Boltzmann constant, and $\tilde{\phi}$ is the self-consistent potential, as dictated by the Poisson equation,

$$\nabla^2\tilde{\phi} = -4\pi\sum_\alpha \delta\tilde{\rho}_\alpha = -4\pi\sum_\alpha q_\alpha\delta\tilde{n}_\alpha, \tag{8}$$

where $\delta\tilde{\rho}_\alpha = \delta\tilde{\rho}_\alpha(\mathbf{r}, t) = q_\alpha\delta\tilde{n}_\alpha(\mathbf{r}, t)$ is the field-induced change in the charge density of plasma particles of sort $\alpha$. $\gamma_\alpha$ in Eq. (6) is the coefficient that fluid-dynamics theories of stimulated light scattering in plasmas take equal to $\gamma_e = 1$ for electrons and $\gamma_e = 3$ [15 – 17] and whose physical content will be discussed below in this paper.

One of the central goals of our study is to explore whether moderate-intensity regimes of stimulated scattering can be understood in terms of constitutive relations that express the material response to an electromagnetic driver field via a power-series expansion of material polarization, with nonlinear-optical susceptibilities $\chi^{(n)}$ defined as field-independent coefficients in such an expansion. With such constitutive relations in place, the field-evolution equations will connect the SBS and SRS gain $g(\Omega)$ of the Stokes field with frequency $\omega_1 = \omega_0 - \Omega$ induced by a pump with frequency $\omega_0$ [Eq. (1)] to the imaginary part of the Fourier-space third-order nonlinear susceptibility $\chi_s^{(3)}(\Omega) = \chi^{(3)}(\omega_1; \omega_0, -\omega_0, \omega_1)$. Such a framework would lay grounds for a unified treatment of stimulated light scattering across different areas of optical physics from fiber optics and neutral gases to laser – plasma interactions. As a part of this analysis, we also aim to resolve the limits of nonlinear-susceptibility-based approach in strong-field theories of stimulated light scattering in plasmas.

To reach these goals, we seek to describe moderate-intensity regimes of stimulated light scattering in plasmas in terms of constitutive relations, expressing the third-order field-induced polarization $\mathbf{P}_{\mathrm{p}}^{(3)}$ in plasmas as a trilinear form of the driver field. Specifically, for the above-outlined archetypal setting of stimulated scattering of the laser field (1) by a plasma wave with $\Omega = \omega_0 - \omega_1$ and $\mathbf{q} = \mathbf{k}_0 - \mathbf{k}_1$, we aim to establish a constitutive relation

$$\mathbf{P}_{\mathrm{p}}^{(3)}(\Omega) = \chi_p^{(3)}(\Omega)\mathbf{E}_0(\mathbf{E}_0^* \cdot \mathbf{E}_1), \qquad (9)$$

expressing the nonlinear polarization at the frequency $\Omega$ in terms of a suitable third-order susceptibility $\chi_p^{(3)}$ and the Fourier amplitudes of the pertinent electric fields.

## 3. Electron and ion susceptibilities for collisionless laser – plasma interactions

Because the nonlinear-optical response of plasmas is a function of collective plasma properties rather than properties that could be related to local field-induced dipoles, an adequate analysis of plasma nonlinear optics should solve the field-evolution equations for the laser driver jointly with suitable kinetic equations for the probability distribution functions of plasma particles [2 – 5, 15, 85, 95, 96]. In the collisionless limit, i.e., in the regime where all the mean free paths $\ell_{\alpha\beta}$ for the collisions between plasma particles $\alpha$ and $\beta$ satisfy $q\ell_{\alpha\beta} \gg 1$, the probability distribution function $f_\alpha = f_\alpha(\mathbf{r}, \mathbf{v}, t)$ of plasma particles of sort $\alpha$ with mass $m_\alpha$ in the position – velocity $(\mathbf{r}, \mathbf{v})$ space, is found as a solution to the collisionless Boltzmann equation

$$\frac{\partial f_\alpha}{\partial t} + \mathbf{v} \cdot \nabla f_\alpha + \frac{\mathbf{G}_\alpha}{m_\alpha} \cdot \frac{\partial f_\alpha}{\partial \mathbf{v}} = 0 \quad , \qquad (10)$$

where $\mathbf{G}_\alpha$ is the force acting on plasma particles of sort $\alpha$.

With the driver force

$$\mathbf{G}_\alpha = q_\alpha \mathbf{E} + q_\alpha \frac{\mathbf{v}}{c} \times \mathbf{B} \quad , \qquad (11)$$

defined in terms of the self-consistent fields $\mathbf{E}$ and $\mathbf{B}$ acting on plasma particles with charge $q_\alpha$, Eq. (10) becomes the Vlasov equation [133 – 138].

Pertaining to stimulated light scattering, however, is not the $\mathbf{G}_\alpha$ force in Eq. (11), but the ponderomotive force $\mathbf{F}_\alpha(\mathbf{r},t)$ as defined by in Eqs. (2) – (4). Because ions are much heavier than electrons, $m_i \gg m_e$, this force drives electrons, but has virtually no effect on ions, $|U_i| \ll |U_e|$, thus giving rise to a potential

$$\tilde{\phi}_\alpha(\mathbf{r},t) = \tilde{\phi}(\mathbf{r},t) + \tilde{U}_\alpha(\mathbf{r},t)/q_\alpha, \qquad (12)$$

with self-consistent potential $\tilde{\phi}(\mathbf{r},t)$ as defined by Eq. (8). Since $|U_i| \ll |U_e|$, Eq. (12) yields $\tilde{\phi}_e(\mathbf{r},t) = \tilde{\phi}(\mathbf{r},t) + \tilde{U}_e(\mathbf{r},t)/q_e$ for electrons and $\tilde{\phi}_i(\mathbf{r},t) \approx \tilde{\phi}(\mathbf{r},t)$ for ions.

As a standard linearization procedure, we search for the solution to Eq. (10) in the form of an expansion

$$f_\alpha(\mathbf{r},\mathbf{v},t) = f_{\alpha 0}(\mathbf{v}) + \delta f_\alpha(\mathbf{r},\mathbf{v},t) \qquad (13)$$

about the equilibrium stationary zero-field distribution $f_{\alpha 0}(\mathbf{r},\mathbf{v})$ and a small perturbation $\delta f_\alpha(\mathbf{r},\mathbf{v},t)$ due to the external force. With the external force $\mathbf{G}_\alpha(\mathbf{r},t) = \mathbf{F}_\alpha(\mathbf{r},t)$ as defined by Eqs. (2) – (4) and with $\delta f_\alpha(\mathbf{r},\mathbf{v},t) = \delta f_\alpha(\mathbf{q},\mathbf{v},\Omega)\exp[i(\mathbf{q}\cdot\mathbf{r} - \Omega t)]$, $\tilde{\phi}_\alpha(\mathbf{r},t) = \phi_\alpha \exp[i(\mathbf{q}\cdot\mathbf{r} - \Omega t)]$, and $\tilde{\phi}(\mathbf{r},t) = \phi \exp[i(\mathbf{q}\cdot\mathbf{r} - \Omega t)]$, the solution to the linearized collisionless kinetic equation is

$$\delta f_\alpha(\mathbf{q},\mathbf{v},\Omega) = \frac{q_\alpha \phi_\alpha}{m_\alpha(\mathbf{q}\cdot\mathbf{v} - \Omega)}\mathbf{q}\cdot\frac{\partial f_{\alpha 0}}{\partial \mathbf{v}} \quad , \qquad (14)$$

where $\phi_\alpha = \phi + U_\alpha/q_\alpha$.

The changes in plasma particle densities $\delta\tilde{n}_\alpha(\mathbf{r},t) = \delta n_\alpha \exp[i(\mathbf{q}\cdot\mathbf{r} - \Omega t)]$ induced by the ponderomotive force are found by integrating Eq. (14) over the velocity space,

$$\delta n_\alpha = \frac{q_\alpha}{m_e}\phi_\alpha \int \mathbf{q}\cdot\frac{\partial f_{\alpha 0}}{\partial \mathbf{v}}\frac{d^3 v}{\mathbf{q}\cdot\mathbf{v} - \Omega} \quad . \qquad (15)$$

The linearity of $\delta n_\alpha$ as a function of the potential $\phi_\alpha$ in Eq. (15) is guaranteed by the linearization (13) of the kinetic equation (10), which searches for the correction $\delta f_\alpha(\mathbf{r},\mathbf{v},t)$ to the zero-field distribution $f_{\alpha 0}(\mathbf{v})$ in the first order in the external force. The susceptibilities of plasma particles of sort $\alpha$ are read off the solution for $\delta n_\alpha$ in Eq. (15) as

$$\chi_\alpha^{\mathrm{kt}} = \frac{4\pi q_\alpha^2}{q^2 m_\alpha}\int \mathbf{q}\cdot\frac{\partial f_{\alpha 0}}{\partial \mathbf{v}}\frac{d^3 v}{\Omega - \mathbf{q}\cdot\mathbf{v}} \quad . \qquad (16)$$

Derivation of these solutions for plasma susceptibilities from the collisionless kinetic equation (10) is provided in the earlier work on SBS and SRS in plasmas [2 – 5, 15, 85, 95, 96]. Combining Eqs. (12), (15), and (16) with the Poisson equation [Eq. (8)] leads to

$$\delta n_e/n_{e0} = -\left[q^2c^2/\left(2\omega_{pe}^2\right)\right]\mathcal{K}\mathbf{a}_0 \cdot \mathbf{a}_1^*, \tag{17}$$

with $\mathbf{a}_j = [e/(m_e c^2)]\mathbf{A}_j$ and

$$\mathcal{K}_{\mathrm{kt}} = \frac{\chi_e^{\mathrm{kt}}\left(1+\chi_i^{\mathrm{kt}}\right)}{1+\chi_e^{\mathrm{kt}}+\chi_i^{\mathrm{kt}}} \quad . \tag{18}$$

Eq. (18) provides powerful insights into stimulated Brillouin scattering in plasmas, shedding light on the properties of SBS in a broad range of laser – plasma interaction settings that can be adequately described in terms of collisionless kinetic equations [2 – 5, 15, 85, 95, 96]. In the following section, we will show that the realm of the algebra of $\mathcal{K}$, whereby $\chi_e$ and $\chi_i$ are mixed into a self-consistent plasma response to an external field, extends way beyond the kinetic-theory models of stimulated light scattering, providing a framework for the analysis of stimulated light scattering in a vast variety of laser – plasma interaction settings.

## 4. Nonlinear susceptibility from self-consistent treatment

To lay out the general procedure for the derivation of nonlinear susceptibilities for stimulated light scattering in plasmas, we consider the ponderomotive force [Eqs. (2) – (4)] induced by a laser field (1) and search for $\tilde{\phi}(\mathbf{r},t)$ and $\delta\tilde{\rho}_\alpha(\mathbf{r},t)$ in the form of $\tilde{\phi}(\mathbf{r},t) = \phi\exp[i(\mathbf{q}\cdot\mathbf{r} - \Omega t)]$ and $\delta\tilde{\rho}_\alpha(\mathbf{r},t) = \delta\rho_\alpha\exp[i(\mathbf{q}\cdot\mathbf{r} - \Omega t)]$. For a plasma consisting of electrons and ions of one sort, the Poisson equation [Eq. (8)] gives

$$q^2\phi = 4\pi(\delta\rho_e + \delta\rho_i). \tag{19}$$

We now invoke a standard definition of linear susceptibilities [99 – 102] as proportionality coefficients connecting polarization induced in a medium to the electric field that induces this polarization. Susceptibilities of plasma particles $\chi_\alpha$ consistent with this definition should be found as coefficients in the relation between polarization $\widetilde{\mathbf{P}}_\alpha(\mathbf{r},t)$ originating from the electric current of plasma particles $\alpha$, $\mathbf{j}_\alpha = \partial\widetilde{\mathbf{P}}_\alpha/\partial t$, and the field $\widetilde{\boldsymbol{\mathcal{E}}}_\alpha(\mathbf{r},t)$ that drives this current. In Fourier space,

$$\mathbf{P}_\alpha(\omega) = \frac{\chi_\alpha(\omega)}{4\pi}\boldsymbol{\mathcal{E}}_\alpha(\omega) \quad . \tag{20}$$

In self-consistent treatment, the fields $\boldsymbol{\mathcal{E}}_\alpha(\omega)$ in Eq. (20) are defined via a Fourier transform of $\widetilde{\boldsymbol{\mathcal{E}}}_\alpha(\mathbf{r},t) = -\nabla\tilde{\phi}_\alpha(\mathbf{r},t)$, leading to $\boldsymbol{\mathcal{E}}_\alpha = -i\mathbf{q}\phi_\alpha$. Because ions are much heavier than electrons, $m_i \gg m_e$, the ponderomotive force [Eqs. (2) – (4)] has virtually no effect on ions, $|U_i| \ll |U_e|$, so that $\boldsymbol{\mathcal{E}}_e = -i\mathbf{q}(\phi - U_e/e)$ and $\boldsymbol{\mathcal{E}}_i \approx -i\mathbf{q}\phi$.

The electric current $\mathbf{j}_\alpha$, on the other hand, connects to the charge density $\delta\tilde{\rho}_\alpha(\mathbf{r},t)$ via the continuity equation expressed as

$$\partial\delta\tilde{\rho}_\alpha/\partial t + \nabla\cdot\mathbf{j}_\alpha = 0. \tag{21}$$

Combining Eq. (21) with $\mathbf{j}_\alpha = \partial\widetilde{\mathbf{P}}_\alpha/\partial t$ gives

$$\delta\tilde{\rho}_\alpha(\mathbf{r},t) = -\nabla\cdot\widetilde{\mathbf{P}}_\alpha(\mathbf{r},t). \tag{22}$$

Fourier transform of Eq. (22) yields $\delta\rho_\alpha = -i\mathbf{q}\cdot\mathbf{P}_\alpha$. Expressing $\mathbf{P}_\alpha$ in this relation in terms of the fields $\boldsymbol{\mathcal{E}}_\alpha$ (or the potential $\phi_\alpha = iq^{-1}\mathcal{E}_\alpha$), we find

$$\delta\rho_e = (4\pi)^{-1}q^2\chi_e\phi_e = (4\pi)^{-1}q^2\chi_e(U_e/e - \phi) \tag{23}$$

for electrons and

$$\delta\rho_i = -(4\pi)^{-1}q^2\chi_i\phi \tag{24}$$

for ions.

Combining Eqs. (23) and (24) with the Poisson equation (19), we express the field-induced change in the electron charge density in terms of $U_e$ as

$$\delta\rho_e = (4\pi e)^{-1}q^2\mathcal{K}U_e, \tag{25}$$

where

$$\mathcal{K} = \frac{\chi_e(1+\chi_i)}{1+\chi_e+\chi_i} \quad . \tag{26}$$

With $\chi_e = \chi_e^{\mathrm{kt}}$ and $\chi_i = \chi_i^{\mathrm{kt}}$ , where $\chi_e^{\mathrm{kt}}$ and $\chi_i^{\mathrm{kt}}$ are electron and ion susceptibilities as defined by Eq. (16), the solution for $\mathcal{K}$ found in Eq. (26) reduces to $\mathcal{K}_{\mathrm{kt}}$ in Eq. (18), derived for a plasma described by the collisionless kinetic equation. We now see, from Eqs. (19) – (26), that, while the individual susceptibilities of plasma constituents $\chi_e$ and $\chi_i$ are functions of plasma parameters, depending on the specific regime of laser – plasma interactions, the realm of the algebra of $\mathcal{K}$, whereby $\chi_e$ and $\chi_i$ are mixed into a self-consistent plasma response to an external field, extends way beyond the kinetic-theory models of stimulated light scattering, providing a framework for the analysis of stimulated light scattering in a vast variety of laser – plasma interaction settings.

To gain better insights into the physical content of $\mathcal{K}$, we observe that, while the susceptibilities $\chi_e$ and $\chi_i$ are defined [Eqs. (23), (24)] via the linear response of electrons and ions to the self-consistent plasma fields $\boldsymbol{\mathcal{E}}_e$ and $\boldsymbol{\mathcal{E}}_i$, acting on electrons and ions, $\mathcal{K}$ is found [Eq. (26)] from the electron linear response to the external force that originates from the ponderomotive force $\mathbf{F}_\alpha$ and that emerges as a part of $\mathbf{F}_\alpha$-induced collective plasma dynamics, in which the electron and ion plasma constituents are coupled by the electrostatic interaction as expressed by the Poisson equation (8) for the charge densities $\delta\tilde{\rho}_\alpha$ and the self-consistent potential $\tilde{\phi}(\mathbf{r},t)$.

Recognizing

$$\varepsilon_p = 1 + \chi_e + \chi_i \tag{27}$$

as plasma dielectric function, we can express $\mathcal{K}$ as the sum

$$\mathcal{K} = \chi_e - \chi_e^2/\varepsilon_p \tag{28}$$

of the individual electron susceptibility $\chi_e$ and the collective-response term $\chi_e^2/\varepsilon_p$, which describes the collective plasma response with resonances at frequencies of plasma eigenmodes, found from

$$\varepsilon_p(\Omega) = 1 + \chi_e + \chi_i = 0. \tag{29}$$

The laser-induced change in the electron charge density found in Eq. (25) translates into the electron-density change

$$\delta n_e = (4\pi)^{-1}(q/e)^2\mathcal{K}U_e, \tag{30}$$

giving rise to an electric current with a current density

$$\mathbf{j}_e = i\frac{e^2\delta n_e^*}{2m_e\omega_0}\mathbf{E}_0 \tag{31}$$

and inducing nonlinear polarization

$$\mathbf{P}_e = \frac{i}{\omega_1}\mathbf{j}_e = -\frac{e^2\delta n_e^*}{2m_e\omega_0\omega_1}\mathbf{E}_0 \quad . \tag{32}$$

Combining Eqs. (4), (30), and (32) reveals the formal structure of $\mathbf{P}_e$ in Eq. (32) as a trilinear form of the laser field,

$$\mathbf{P}_e = \mathbf{P}_\mathrm{p}^{(3)}(\Omega) = \frac{e^2q^2}{16\pi m_e^2\omega_0^2\omega_1^2}\mathcal{K}^*(\mathbf{q},\Omega)\mathbf{E}_0(\mathbf{E}_1\cdot\mathbf{E}_0^*) \quad . \tag{33}$$

$\mathbf{P}_e$ is thus recognized as the sought-for third-order polarization $\mathbf{P}_\mathrm{p}^{(3)}(\Omega)$ [cf. Eqs. (9) and (33)], with cubic susceptibility

$$\chi_p^{(3)} = \frac{e^2q^2}{16\pi m_e^2\omega_0^2\omega_1^2}\mathcal{K}^*(\mathbf{q},\Omega) \quad . \tag{34}$$

To understand the physical content of the nonlinear susceptibility as expressed by Eq. (34), we represent it as a product

$$\chi_p^{(3)} = c_1c_2(-c_1), \tag{35}$$

with

$$c_1 = \frac{e^2}{2m_e\omega_0\omega_1} \quad , \tag{36}$$

and

$$c_2 = -(4\pi)^{-1}(q/e)^2\mathcal{K}^*. \tag{37}$$

The first multiplier in Eq. (35) quantifies the efficiency of conversion from the laser-driven beat wave to the ponderomotive potential of electrons. Indeed, electrons driven by the pump and Stokes fields acquire quiver velocities $\mathbf{v}_{0,1} = -ie\mathbf{E}_{0,1}/(m_e\omega_{0,1})$, giving rise to the beat ponderomotive potential energy [Eqs. (3), (4)] via the interference term in the quiver kinetic energy, $m_e v_0 v_1^*/2 = [e^2/(2m_e\omega_0\omega_1)]E_0E_1^* = c_1E_0E_1^*$. As is readily seen, $m_e v_0 v_1^*/2 = c_1E_0E_1^*$ with $c_1$ as defined by Eq. (36), recovers $U_e$ in Eq. (4).

The second factor in Eq. (35), $c_2$, picks up where $c_1$ has left off as it defines the efficiency of the process whereby the ponderomotive force induces electron density modulation. In its capacity as the $U_e$-to-$\delta n_e$ conversion efficiency, $c_2$ is read off directly from Eq. (30).

Finally, the third multiplier in Eq. (35), $-c_1$, is the descriptor of conversion from electron density modulation to polarization $\mathbf{P}_\mathrm{p}^{(3)}(\Omega)$. Indeed, in the presence of the pump field, electron density modulation $\delta n_e^*$ in the conjugate electron-density grating gives rise to the electron current density at the Stokes frequency as described by Eq. (31), inducing nonlinear polarization $\mathbf{P}_\mathrm{p}^{(3)}(\Omega)$ as prescribed by Eq. (32).

## 5. Nonlinear refraction and stimulated gain

The standard model of stimulated light scattering in plasmas as expressed by Eqs. (2) – (8) does not offer a simple connection to the nonlinear polarization $\mathbf{P}_\mathrm{p}^{(3)}(\Omega)$ as defined by Eq. (32). In this model, stimulated light scattering is treated as a parametric instability process, in which the pump field with frequency $\omega_0$ and wave vector $\mathbf{k}_0$ decays into a Stokes electromagnetic wave with frequency $\omega_1$ and wave vector $\mathbf{k}_1$ and a plasma wave, e.g., the IAW in SBS and the electron Langmuir wave in SRS, with frequency $\Omega = \omega_0 - \omega_1$ and wave vector $\mathbf{q} = \mathbf{k}_0 - \mathbf{k}_1$. To understand the status of the nonlinear polarization $\mathbf{P}_\mathrm{p}^{(3)}(\Omega)$ in the model of Eqs. (2) – (8), we express the nonlinear source term on the right-hand side of Eq. (5) in terms of $\delta n_e$ as defined in Eq. (30) and isolate the equation for the amplitude of the Stokes field as

$$[D_s(k_1,\omega_1)+(q^2c^2/4)\mathcal{K}^*(q,\Omega)|\mathbf{a}_0|^2]\mathbf{a}_1=0, \quad (38)$$

where

$$D_s(k_1,\omega_1)=\omega_1^2-c^2k_1^2-\omega_{pe}^2. \quad (39)$$

Eq. (38) is not equivalent to the set of equations (5) – (8). Indeed, $\delta\tilde{n}_e$ in the nonlinear source term on the right-hand side of Eq. (5) is a function of position and time, $\delta\tilde{n}_e=\delta\tilde{n}_e(\mathbf{r},t)$, found from the joint solution to Eqs. (5) – (8). Such $\delta\tilde{n}_e(\mathbf{r},t)$ can be replaced by $\delta n_e\exp[i(\mathbf{q}\cdot\mathbf{r}-\Omega t)]$ with position-independent amplitude $\delta n_e$ only when transient processes are negligible. This regime is only attainable when the buildup time of the field-induced electron density grating is much shorter than the laser pulse.

Operating in the weak-field limit, where all the effects induced by the laser field (1) can be treated perturbatively, we express $k_1$ in Eqs. (38) and (39) as a sum $k_1=k_{10}+\Delta k_1$, isolating the field-induced correction $\Delta k_1$ from the zero-field Stokes wavenumber $k_{10}=\omega_1\eta_{10}/c$, with zero-field refractive index $\eta_{10}$, meeting the dispersion relation for the Stokes wave, $D_s(k_{10},\omega_1)=0$. Expanding $D_s(k_1,\omega_1)$ as a power series in small $\Delta k_1$ about $k_{10}$,

$$D_s(k_1,\omega_1)\approx -2c^2|k_{10}|\Delta k_1, \quad (40)$$

and solving Eq. (38) for $\Delta k_1$, we find

$$\Delta k_1=[q^2/(8|k_{10}|)]\mathcal{K}^*(q,\Omega)|\mathbf{a}_0|^2. \quad (41)$$

The refractive-index change that the laser pump induces at the Stokes frequency as a part of this laser – plasma interaction is thus given by

$$\Delta\eta=\frac{e^2q^2}{8\eta_{10}m_e^2\omega_0^2\omega_1^2}\mathcal{K}^*(q,\Omega)|\mathbf{E}_0|^2 \quad . \quad (42)$$

The refractive index change $\Delta\eta$ in Eq. (42) can be expressed as $\Delta\eta=\kappa|\mathbf{E}_0|^2$, with

$$\kappa=\frac{2\pi}{\eta_{10}}\chi_p^{(3)} \quad . \quad (43)$$

The real part of $\kappa$ as defined by Eq. (43) plays the role of the nonlinear coupling coefficient similar to the nonlinear refractive index $n_2$ in optical Kerr effect [100 – 102], connecting the field-induced refractive index change $\Delta\eta$ to the intensity of the laser field that gives rise to this change. Unlike the Kerr-effect $n_2$, however, $\kappa$ in Eq. (43) is complex, with its imaginary part translating into the gain that the Stokes field is subject to as a part of stimulated light scattering.

The real parts of $\Delta k_1$ is thus read out by the Stokes field via the nonlinear, pump-intensity-dependent phase $\Phi_1 = \mathrm{Re}(\Delta k_1)z$ that this field acquires as it propagates along the $z$-axis. The imaginary part of $\Delta k_1$, on the other hand, translates into the gain or loss of the Stokes field, giving rise to an exponential attenuation or exponential buildup of the Stokes intensity,

$$I_s(z) = I_{s0} \exp[-2\mathrm{Im}(\Delta k_1)z] = I_{s0} \exp[g_s|\mathbf{E}_0|^2 z], \quad (44)$$

with

$$g_s = -\frac{4\pi}{\eta_{10}}\frac{\omega_1}{c}\mathrm{Im}\left[\chi_p^{(3)}\right] \quad . \quad (45)$$

Because in SBS $\Omega \ll \omega_0, \omega_1$, the grating wavenumber $q = \{(k_0 - k_1)^2 + 4k_0 k_1[\sin(\psi/2)]^2\}^{1/2}$ is accurately approximated as $q \approx 2k_0 \sin(\psi/2)$ for all $\psi$ larger than, typically, a few hundredths of a degree. With such an approximation, Eqs. (44) and (45) become

$$\Delta\eta \approx (\eta_{10}/2)\mathcal{K}^*(\mathbf{q},\Omega)|\mathbf{a}_0|^2[\sin(\psi/2)]^2 \quad (46)$$

and

$$\mathcal{G}_s = g_s|\mathbf{E}_0|^2 = \eta_{10}\mathrm{Im}(\mathcal{K})|\mathbf{a}_0|^2[\sin(\psi/2)]^2. \quad (47)$$

Eqs. (41) – (47) are manifestly different in their scope and physical content from Eqs. (5) – (8). Indeed, while Eqs. (5) – (8) express the general physical model that describes stimulated light scattering as a $\omega_0 \to \omega_1 + \Omega$, $\mathbf{k}_0 \to \mathbf{k}_1 + \mathbf{q}$ decay instability of the pump field, Eqs. (41) – (47) are derived as the weak-field, steady-state limit of Eqs. (5) – (8). Eqs. (41) – (47) connect plasma SBS and SRS to the third-order polarization and express the SBS/SRS gain in terms of the imaginary part of $\chi_p^{(3)}$, thus recovering the signature result of SBS and SRS theories for neutral gases, liquids and solids [100 – 102] and laying grounds for a unified treatment of stimulated light scattering across a vast area of optical physics, spanning from fiber optics and neutral gases to laser – plasma physics.

## 6. $\chi_p^{(3)}$ from plasma fluid dynamics

Derivation of Eqs. (26), (43) – (47) from Eqs. (12), (19) – (24) shows that the algebra whereby $\chi_e$ and $\chi_i$ are mixed into the nonlinear susceptibility $\chi_p^{(3)}$ [Eqs. (26), (34)] and, hence, into the nonlinear refractive index [Eq. (43)] and stimulated gain $g_s$ [Eqs. (45), (47)] is guaranteed by the Poisson equation [Eqs. (8)] and continuity equation [Eq. (7)] linearized with respect to $\delta\tilde{n}_\alpha$. Because the form of both equations is the same for plasmas obeying the kinetic description of Eqs. (10), (13) and plasmas with fluid dynamics as defined by Eqs. (5) – (8), the algebra of $\chi_e$ and $\chi_i$ mixing into $\chi_p^{(3)}$ and $g_s$ is common to the nonlinear response of both classes of plasma. While the algebra of $\chi_\alpha$ mixing into $\chi_p^{(3)}$ remains as prescribed by Eqs. (26), (34), individual susceptibilities $\chi_\alpha$ of plasma particles are not immune to changes in the way the collective plasma response sets in as a part of plasma dynamics. These susceptibilities need to be defined for a specific class of laser – plasma interactions.

In the kinetic-equation framework [Eqs. (2) – (4), (10)], linearization of the kinetic equation via Eq. (13) leads (Section 4) to the solutions for electron and ion susceptibilities in the form of Eqs. (16). Derivation of these solutions for $\chi_\alpha$ does not require any additional approximations beyond the approximations leading up to Eq. (14). In the plasma-fluid framework, however, this set of approximations is insufficient to derive closed-form solutions for $\chi_\alpha$.

To appreciate the physical content of susceptibilities $\chi_\alpha$ in plasma fluid dynamics as dictated by the general framework of Eqs. (26), (33) – (37), we observe that the key quantifiers of plasma fluid dynamics are found as the respective moments of the distribution functions $f_\alpha(\mathbf{r},\mathbf{v},t)$. Specifically, the number densities $n_\alpha(\mathbf{r},t)$ and velocities $\tilde{\mathbf{u}}_\alpha(\mathbf{r},t)$ of plasma particles are given by respectively the zeroth and first moments of the probability densities $f_\alpha(\mathbf{r},\mathbf{v},t)$,

$$n_\alpha(\mathbf{r},t) = \int f_\alpha(\mathbf{r},\mathbf{v},t) d^3v, \qquad (48)$$

$$\tilde{\mathbf{u}}_\alpha(\mathbf{r},t) = n_\alpha^{-1} \int \mathbf{v} f_\alpha(\mathbf{r},\mathbf{v},t) d^3v. \qquad (49)$$

Integrating the kinetic equation (10) over the velocity space with $f_\alpha(\mathbf{r},\mathbf{v},t) \to 0$ for $v \to \infty$ yields the continuity equation

$$\partial n_\alpha / \partial t + \nabla \cdot (n_\alpha \tilde{\mathbf{u}}_\alpha) = 0. \qquad (50)$$

Linearization of Eq. (50) via $n_\alpha = n_{\alpha 0} + \delta\tilde{n}_\alpha$ for $|\delta\tilde{n}_\alpha| \ll n_{\alpha 0}$ recovers the continuity equation (7) in the fluid-dynamics framework of Eqs (5) – (8). In the Fourier space, with $\tilde{\mathbf{u}}_\alpha(\mathbf{r},t) = \mathbf{u}_\alpha \exp[i(\mathbf{q}\cdot\mathbf{r} - \Omega t)]$, this equation gives

$$-i\Omega\delta n_\alpha + i n_{\alpha 0} \mathbf{q}\cdot\mathbf{u}_\alpha = 0. \qquad (51)$$

Multiplying the continuity equation (50) by $q_\alpha$, recognizing $q_\alpha n_\alpha \mathbf{u}_\alpha$ in the second term as polarization current density, $\mathbf{j}_\alpha = q_\alpha n_\alpha \mathbf{u}_\alpha$, and expressing this term via polarization $\tilde{\mathbf{P}}_\alpha(\mathbf{r},t)$ induced by the current $\mathbf{j}_\alpha$, $\mathbf{j}_\alpha = \partial\tilde{\mathbf{P}}_\alpha/\partial t$, we recover Eq. (22), relating $\delta\tilde{\rho}_\alpha(\mathbf{r},t)$ to $\nabla\cdot\tilde{\mathbf{P}}_\alpha(\mathbf{r},t)$.

To define susceptibilities $\chi_\alpha$ via the self-consistent procedure as described by Eqs. (20) – (24), we need to express the field-induced charge-density change $\delta\rho_\alpha$ as a linear function of $\phi_\alpha$. As a step toward this goal, the Fourier-space linearized continuity equation (51) allows $\delta n_\alpha$ to be expressed as a linear function of $\mathbf{u}_\alpha$. The equation for $\mathbf{u}_\alpha$, on the other hand, is obtained by taking the first moment of the kinetic equation (10). Indeed, multiplying Eq. (10) by $m_\alpha\mathbf{v}$ and integrating the resulting equation over the velocity space, we find

$$m_\alpha \mathcal{n}_\alpha(\partial\tilde{\mathbf{u}}_\alpha/\partial t + \tilde{\mathbf{u}}_\alpha\nabla\tilde{\mathbf{u}}_\alpha) = -\mathcal{q}_\alpha \mathcal{n}_\alpha\nabla\tilde{\phi}_\alpha - \nabla\cdot\boldsymbol{\mathcal{P}}_\alpha, \quad (52)$$

where $\boldsymbol{\mathcal{P}}_\alpha$ is the pressure tensor, defined as the pressure moment of $f_\alpha(\mathbf{r},\mathbf{v},t)$, with its tensor components given by

$$\mathcal{P}_{\alpha,ij} = m_\alpha \int \left(v_i - \tilde{u}_{\alpha,i}\right)\left(v_j - \tilde{u}_{\alpha,i}\right) f_\alpha(\mathbf{r},\mathbf{v},t) d^3v. \quad (53)$$

For isotropic pressure, $\mathcal{P}_{\alpha,ij} = \wp_\alpha\delta_{ij}$, Eq. (52) becomes

$$m_\alpha \mathcal{n}_\alpha(\partial\tilde{\mathbf{u}}_\alpha/\partial t + \tilde{\mathbf{u}}_\alpha\nabla\tilde{\mathbf{u}}_\alpha) = -\mathcal{q}_\alpha \mathcal{n}_\alpha\nabla\tilde{\phi}_\alpha - \nabla\wp_\alpha. \quad (54)$$

Isolating the longitudinal component of $\tilde{\mathbf{u}}_\alpha$ along $\mathbf{q}$, $\tilde{u}_\alpha(\mathbf{r},t) = \mathcal{n}_\alpha^{-1}\int v_q f_\alpha(\mathbf{r},\mathbf{v},t)d^3v$, and linearizing Eq. (54) via $\mathcal{n}_\alpha = n_{\alpha 0} + \delta\tilde{n}_\alpha$ and $\wp_\alpha = \wp_{\alpha 0} + \delta\tilde{\wp}_\alpha$ with $|\delta\tilde{n}_\alpha| \ll n_{\alpha 0}$ and $|\delta\tilde{\wp}_\alpha| \ll \wp_{\alpha 0}$, we find in the Fourier space with $\delta\tilde{u}_\alpha(\mathbf{r},t) = \delta u_\alpha \exp[i(\mathbf{q}\cdot\mathbf{r} - \Omega t)]$ and $\delta\tilde{\wp}_\alpha(\mathbf{r},t) = \delta\wp_\alpha \exp[i(\mathbf{q}\cdot\mathbf{r} - \Omega t)]$

$$-i\Omega m_\alpha n_{\alpha 0} u_\alpha + iq n_{\alpha 0}\mathcal{q}_\alpha\phi_\alpha + iq\delta\wp_\alpha = 0. \quad (55)$$

Combining Eq. (55) with the continuity equation (51) yields

$$m_\alpha\Omega^2\delta n_\alpha = q^2 n_{\alpha 0}\mathcal{q}_\alpha\phi_\alpha + q^2\delta\wp_\alpha. \quad (56)$$

We see from Eqs. (52) – (56) that, while the first moment of the kinetic equation (10) does provide an equation for $u_\alpha$ [Eq. (52)], the set of equations for $\delta n_\alpha$ is still not closed even with this equation as it involves unknown $\delta\wp_\alpha$, found via the second moment of the distribution function $f_\alpha(\mathbf{r},\mathbf{v},t)$. As a general property of the kinetic equation (10), any system of $l$ evolution equations for $l$ lowest-order moments of the distribution function $f_\alpha(\mathbf{r},\mathbf{v},t)$ is never closed as the evolution equation for its $l$th moment always contains its $(l+1)$th moment. As a manifestation of this general property, the evolution equation for the zeroth-order, density moment of the distribution function $f_\alpha(\mathbf{r},\mathbf{v},t)$ [Eq. (50)] includes its first-order moment – velocity $\tilde{\mathbf{u}}_\alpha$. The evolution equation for $\tilde{\mathbf{u}}_\alpha$, on the other hand, includes the second, pressure moment $\boldsymbol{\mathcal{P}}_\alpha$ [Eq. (52)]. The evolution equation for $\boldsymbol{\mathcal{P}}_\alpha$, in its turn, will contain the third, heat-flux moment. Thus, any system of $l$ evolution equations for $l$ lowest-order moments of the distribution function $f_\alpha(\mathbf{r},\mathbf{v},t)$ can only be resolved if it is truncated for some $l$ by invoking a physically meaningful assumption that would decouple the $l$th moment from the higher moments in the hierarchy of $f_\alpha(\mathbf{r},\mathbf{v},t)$ moments.

Specifically, to be able to express $\delta\rho_\alpha$ as a closed-form linear function of $\phi_\alpha$, as needed for the definition of susceptibilities $\chi_\alpha$ via the general self-consistent procedure as described by Eqs. (20) – (24), we need to set up a physically meaningful closure for a finite set of equations for the lowest-order moments of $f_\alpha(\mathbf{r},\mathbf{v},t)$. A suitable procedure is known in fluid dynamics as polytropic closure [139],

$$\mathcal{p}_\alpha n_\alpha^{-\gamma_\alpha} = \text{const}, \tag{57}$$

with a polytropic index $\gamma_\alpha$.

For linearizable variations of $\mathcal{p}_\alpha$ and $n_\alpha$ as defined above, Eq. (57) gives

$$\delta\mathcal{p}_\alpha = \gamma_\alpha k_B T_\alpha \delta n_\alpha. \tag{58}$$

Substituting Eq. (58) into Eq. (54) recovers the evolution equation for the velocity $\tilde{\mathbf{u}}_\alpha(\mathbf{r},t)$ used as a part of the model of plasma fluid dynamics [Eq. (6)]. On the other hand, substituting Eq. (58) into Eq. (56) and solving the resulting equation for $\delta n_\alpha$, we derive

$$\delta n_\alpha = \frac{n_{\alpha 0} q_\alpha q^2 \phi_\alpha}{m_\alpha \Omega^2 - \gamma_\alpha q^2 k_B T_\alpha} \tag{59}$$

and

$$\delta\rho_\alpha = \frac{n_{\alpha 0} q_\alpha^2 q^2 \phi_\alpha}{m_\alpha \Omega^2 - \gamma_\alpha q^2 k_B T_\alpha} \quad . \tag{60}$$

Eq. (60) provides the sought-for solution expressing $\delta\rho_\alpha$ as a linear function of $\phi_\alpha$. The susceptibilities of plasma particles $\alpha$ are now found via Eqs. (23) and (24), leading to

$$\chi_\alpha^{\text{fl}} = -\frac{\omega_{p\alpha}^2}{\Omega^2 - \gamma_\alpha q^2\, k_B T_\alpha / m_\alpha} \quad , \tag{61}$$

with $\omega_{p\alpha}^2 = 4\pi q_\alpha^2 n_{\alpha 0}/m_\alpha$.

In the cold-fluid limit, i.e., with $T_\alpha \to 0$. Eq. (61) reduces to the well-known approximate expression for the plasma dielectric function

$$\varepsilon_p(\omega) \approx 1 - \sum_\alpha \omega_{p\alpha}^2/\omega^2 \quad . \tag{62}$$

When the motion of ions is negligible, Eq. (62) gives $\varepsilon_p(\omega) \approx 1 - \omega_{pe}^2/\omega^2$.

The electron thermal velocity $v_{Te}$ is much higher than the phase velocity of the ion-acoustic wave. Electron-temperature perturbations are therefore rapidly transported along the IAW and are much smaller than density perturbations, such that, to the leading order, $\delta p_e = k_B T_e \delta n_e + k_B n_e \delta T_e \approx k_B T_e \delta n_e$, dictating the isothermal closure $\gamma_e \approx 1$. For ions, the value of the $\gamma_\alpha$ index in the polytropic closure (57) is chosen in such a way, $\gamma_i = 3$, as to reproduce the equation for small perturbations of the longitudinal pressure $P_{i\parallel} = m_i \int \left(v_x - \tilde{u}_{i,x}\right)^2 f_i d^3 v$ derived by linearizing the second moment of the collisionless kinetic equation (10) with zero heat-flux perturbation [140 – 142]. Such a closure is broadly referred to as the one-dimensional adiabatic closure.

## 7. Ideal-fluid $\chi_p^{(3)}$ vis-à-vis kinetic-theory $\chi_p^{(3)}$

The model of plasma dynamics described by the kinetic equation (10) is distinctly different in its physical content and mathematical structure from the model of fluid plasma dynamics in Eqs. (6) – (8). Yet, both models are broadly used for the analysis of the wide variety of stimulated scattering scenarios in laser – plasma interactions. Specifically, the model of fluid plasma dynamics provides an accurate description of the properties of plasma SBS and SRS in a broad range of laser intensities and pulse widths, including ultrafast stimulated scattering, as well as strong-field SBS and SRS, in which intense laser fields can significantly alter the properties of collective plasma waves, shifting their frequencies relative to the frequencies of Langmuir and IAW plasma-wave eigenmodes.

As is readily seen from Eqs. (16) and (61), the susceptibilities $\chi_\alpha^{\mathrm{kt}}$ and $\chi_\alpha^{\mathrm{fl}}$ that the kinetic theory [Eqs. (10) – (13)] and plasma fluid dynamics [Eqs. (6) – (8)] predict for laser-driven plasma particles are distinctly different in their form. As one of the most important differences, while $\chi_\alpha^{\mathrm{fl}}$ is purely real [Eq. (61)], $\chi_\alpha^{\mathrm{kt}}$ is complex [Eq. (16)]. To understand the physical content of the imaginary part of $\chi_\alpha^{\mathrm{kt}}$, we apply the Sokhotski–Plemelj theorem to the integrand in Eq. (16) to find

$$\mathrm{Im}\chi_\alpha^{\mathrm{kt}} = -\pi \frac{\omega_{p\alpha}^2}{n_{\alpha 0} q^2} \frac{\partial f_{\alpha 0}}{\partial v}\bigg|_{v=q\Omega} . \qquad (63)$$

The imaginary part of $\chi_\alpha^{\mathrm{kt}}$ thus comes entirely from plasma particles whose velocity along **q** is equal to the phase velocity of the beat wave, $v = q\Omega$. For a distribution with $(\partial f_{\alpha 0}/\partial v)|_{v=q\Omega} < 0$, Eq. (63) dictates $\mathrm{Im}\chi_\alpha^{\mathrm{kt}} > 0$, indicating energy transfer from the beat wave to resonant plasma particles, i.e., the Landau damping of the beat wave.

For a Maxwellian zero-field distribution $f_{\alpha 0}(\mathbf{v})$, Eq. (63) gives

$$\mathrm{Im}\chi_\alpha^{\mathrm{kt}} = \frac{\sqrt{\pi}}{q^2\lambda_{D\alpha}^2}\theta_\alpha \exp(-\theta_\alpha^2) \quad , \tag{64}$$

with $\theta_\alpha = \Omega/\left(\sqrt{2}qv_{T\alpha}\right)$.

Eq. (64) recovers the standard result for the Landau damping [121, 136 – 138, 143, 144]. The imaginary part of the nonlinear susceptibility $\chi_p^{(3)}$ calculated with electron and ion susceptibilities $\chi_e^{\mathrm{kt}}$ and $\chi_i^{\mathrm{kt}}$ as dictated by the kinetic theory is thus an expression of Landau damping. Because it is the imaginary part of $\chi_p^{(3)}$ that gives rise to the gain of the Stokes field in stimulated light scattering [Eqs. (45), (47)], the fluid model of plasma dynamics, which, in its canonical form, leads to purely real $\chi_\alpha^{\mathrm{fl}}$ and, hence, purely real $\chi_p^{(3)}$, is insufficient for the description of the perturbative regimes of stimulated light scattering in plasmas.

Yet, this model provides a meaningful benchmark for the analysis of stimulated light scattering in plasmas. Indeed, the kinetic theory provides crucial physical insights into the origin of stimulated light scattering in plasmas, revealing the connection between the stimulated gain and the kinetic properties of laser-driven plasmas, including Landau damping (see Section 3). This approach, however, does not offer equally transparent closed-form solutions for some of the key parameters of stimulated light scattering in plasmas, such as the IAW and Langmuir frequencies. Such solutions are found in the framework of the plasma fluid model, which thus provides the missing physical context by relating stimulated light scattering to plasma dispersion and expressing the IAW and Langmuir frequencies in terms of the parameters of plasma fluid dynamics.

To gain better insights into how the fluid model connects to the kinetic theory of stimulated scattering in plasmas, providing meaningful benchmarks for the kinetic treatment of stimulated scattering, we set the unperturbed, zero-field distribution functions $f_{\alpha 0}(\mathbf{v})$ to be Maxwellian and express the kinetic-theory susceptibilities $\chi_\alpha^{\mathrm{kt}}$ [Eq. (16)] as

$$\chi_\alpha^{\mathrm{kt}} = -\frac{1}{2q^2\lambda_{D\alpha}^2}Z'(\theta_\alpha) = \frac{1}{q^2\lambda_{D\alpha}^2}[1 + \theta_\alpha Z(\theta_\alpha)] \quad , \tag{65}$$

where

$$Z(\xi) = \frac{1}{\sqrt{\pi}} \int \frac{\exp(-t^2)}{t-\xi} dt \qquad (66)$$

is the plasma dispersion function [144 – 146] and $\lambda_{D\alpha} = v_{T\alpha}/\omega_{p\alpha}$ is the Debye length.

Stimulated light scattering in plasmas is enhanced when the frequency $\Omega = \omega_0 - \omega_1$ and the wave vector $\mathbf{q} = \mathbf{k}_0 - \mathbf{k}_1$ of the laser-driven beat wave match the dispersion relation of one of the plasma-wave eigenmodes. In stimulated Brillouin scattering, enabled by the interaction of a laser field of the form of Eq. (1) with low-frequency ion – acoustic waves, the ratio $\Omega/q$ meets inequalities $v_{Ti} \ll \Omega/q \ll v_{Te}$. The properties of electron susceptibilities near the IAW resonance $\Omega \approx \Omega_{\mathrm{IAW}}$ can be therefore understood in terms of $\theta_e \ll 1$ expansions of $\chi_e^{\mathrm{kt}}$ and $\chi_e^{\mathrm{fl}}$,

$$\chi_e^{\mathrm{kt}} \approx \frac{1}{q^2\lambda_{De}^2}\left[1 + i\sqrt{\pi}\theta_e - 2\theta_e^2 - i\sqrt{\pi}\theta_e^3 + \frac{4}{3}\,\theta_e^4 + \ldots\right] \qquad (67)$$

and

$$\chi_e^{\mathrm{fl}} = \frac{1}{q^2\lambda_{De}^2}\frac{1}{1-2\theta_e^2} \approx \frac{1}{q^2\lambda_{De}^2}[1 + 2\theta_e^2 + 4\,\theta_e^4 + \ldots] \quad . \qquad (68)$$

The leading term in both expansions is the same, $\chi_e^{\mathrm{B}} = q^{-2}\lambda_{De}^{-2}$, describing the susceptibility of Boltzmann electrons, i.e., electrons with a Boltzmann distribution $n_e(\mathbf{r},t) = n_{e0}\exp\left[e\tilde{\phi}_\alpha(\mathbf{r},t)/(k_B T_e)\right]$.

The ion susceptibilities near IAW resonances, on the other hand, are found as large-$\theta_i$ expansions of $\chi_i^{\mathrm{kt}}$ and $\chi_i^{\mathrm{fl}}$,

$$\begin{aligned}\chi_i^{\mathrm{kt}} &\approx -\frac{1}{q^2\lambda_{Di}^2}\left[\frac{1}{2\theta_i^2} + \frac{3}{4\theta_i^4} + \frac{15}{8\theta_i^6} + \frac{105}{16\theta_i^8} + \ldots\right] + \frac{i\sqrt{\pi}}{q^2\lambda_{Di}^2}\theta_i\exp(-\theta_i^2) \\ &= -\frac{\omega_{pi}^2}{\Omega^2}\left[1 + 3\left(\frac{qv_{Ti}}{\Omega}\right)^2 + 15\left(\frac{qv_{Ti}}{\Omega}\right)^4 + 105\left(\frac{qv_{Ti}}{\Omega}\right)^6 + \ldots\right] \\ &\quad + \frac{i\sqrt{\pi}}{q^2\lambda_{Di}^2}\theta_i\exp(-\theta_i^2)\end{aligned} \quad , (69)$$

and

$$\begin{aligned}\chi_i^{\mathrm{fl}} &\approx -\frac{1}{q^2\lambda_{Di}^2}\left[\frac{1}{2\theta_i^2} + \frac{3}{4\theta_i^4} + \frac{9}{8\theta_i^6} + \frac{27}{16\theta_i^8} + \ldots\right] \\ &= -\frac{\omega_{pi}^2}{\Omega^2}\left[1 + 3\left(\frac{qv_{Ti}}{\Omega}\right)^2 + 9\left(\frac{qv_{Ti}}{\Omega}\right)^4 + 27\left(\frac{qv_{Ti}}{\Omega}\right)^6 + \ldots\right]\end{aligned} \quad . (70)$$

We observe that the imaginary part of $\chi_i^{\mathrm{kt}}$ in Eq. (69) recovers the exact solution for $\mathrm{Im}\chi_i^{\mathrm{kt}}$ [Eq. (64)] derived by applying the Sokhotski–Plemelj theorem to the general-form kinetic-model solution (16) for $\chi_\alpha^{\mathrm{kt}}$ in plasmas where the initial distributions of particles are Maxwellian.

With the electron susceptibility near the IAW resonance approximated with the leading, Boltzmann-electron term, which is common for $\chi_e^{\mathrm{kt}}$ and $\chi_e^{\mathrm{fl}}$, $\chi_e^{\mathrm{kt,fl}} \approx \chi_e^{\mathrm{B}} = q^{-2}\lambda_{De}^{-2}$, and with ion susceptibility $\chi_i \approx \chi_i^{\mathrm{fl}}$, the dispersion relation of the ion – acoustic wave is expressed as

$$\varepsilon_p(\Omega) = 1 + \frac{1}{q^2\lambda_{De}^2} - \frac{\omega_{pi}^2}{\Omega^2 - 3q^2 v_{Ti}^2} = 0 \quad . \tag{71}$$

Solving Eq. (71) for $\Omega$ yields

$$\Omega_{\mathrm{IAW}}^2 = q^2\left[\frac{3k_BT_i}{m_i} + \frac{Zk_BT_e}{m_i(1+q^2\lambda_{De}^2)}\right] \quad , \tag{72}$$

where $Z$ is the ion charge.

In the limit of $q\lambda_{De} \to 0$, i.e., when the beat-wave wavelength $2\pi/q$ is much larger than the electron Debye length $\lambda_{De}$, i.e., the length scale of charge separation, Eq. (72) reduces to

$$\Omega_{\mathrm{IAW}}^2 \approx \frac{q^2}{m_i}(3k_BT_i + Zk_BT_e) = q^2c_s^2 \quad , \tag{73}$$

where $c_s = [(Zk_BT_e + 3k_BT_i)/m_i]^{1/2}$ is the ion sound speed, with $\Omega_{\mathrm{IAW}} = qc_s$ expressing the standard IAW dispersion relation. As an *a posteriori* verification of inequalities $v_{Ti} \ll \Omega/q \ll v_{Te}$ used to justify power-series expansions of $\chi_\alpha^{\mathrm{kt}}$ and $\chi_\alpha^{\mathrm{fl}}$ in Eqs. (67) – (70), we observe that $c_s^2/v_{Te}^2 = (m_e/m_i)[Z + 3(T_i/T_e)] \ll 1$ and $c_s^2/v_{Ti}^2 = ZT_e/T_i + 3 \gg 1$, as required.

To understand the properties of SBS gain spectra in, we define weakly damped plasma waves as the poles $\Omega_w = \Omega_p - i\Gamma_p$ of the plasma dielectric function $\varepsilon_p(\Omega)$, with $\Omega_p$ being the real plasma-wave frequency, $\mathrm{Re}\left[\varepsilon_p(\Omega_p)\right] = 0$. Expanding $\varepsilon_p(\Omega)$ as a power series about $\Omega_p$, we find $\varepsilon_p(\Omega) \approx \left[\partial(\mathrm{Re}\varepsilon_p)/\partial\Omega\right]_{\Omega_p}(\Omega - \Omega_p) + i\mathrm{Im}\left[\varepsilon_p(\Omega_p)\right]$. The damping constant $\Gamma_p$ is thus

$$\Gamma_p = \mathrm{Im}\left[\varepsilon_p(\Omega_p)\right]/\left[\partial(\mathrm{Re}\varepsilon_p)/\partial\Omega\right]_{\Omega_p} . \tag{74}$$

Provided that the unperturbed distribution functions $f_{\alpha 0}(\mathbf{v})$ of all plasma particles are Maxwellian, we find

$$\mathrm{Im}\varepsilon_p = \sqrt{\pi}q^{-2}\sum_\alpha \lambda_{D\alpha}^{-2}\theta_\alpha \exp(-\theta_\alpha^2) \tag{75}$$

and

$$\partial(\mathrm{Re}\varepsilon_p)/\partial\Omega = \frac{1}{\sqrt{2}q^3}\sum\nolimits_\alpha \frac{1}{\lambda_{D\alpha}^2 v_{T\alpha}}[(4\theta_\alpha^2 - 2)\mathcal{F}(\theta_\alpha) - 2\theta_\alpha] \quad , (76)$$

where $\mathcal{F}(\xi)$ is the Dawson function.

For low ion temperatures, $T_i \ll ZT_e$, Eqs. (75) and (76) offer an insightful estimate

$$\Gamma_p \approx \Omega_p\sqrt{\frac{\pi}{8}}\left[\left(Z\frac{m_e}{m_i}\right)^{1/2} + \left(Z\frac{T_e}{T_i}\right)^{3/2}\exp\left(-\frac{ZT_e}{2T_i}\right)\right] \quad . \quad (77)$$

For Langmuir waves, providing a source of stimulated Raman scattering in plasmas, $\Omega/q \gg v_{Te}$. The motion of ions is negligible for such waves, and ions can be treated as stationary, $\chi_i \approx 0$. To understand the properties of the electron susceptibility near the Langmuir resonances, we expand $\mathrm{Re}\chi_e^{\mathrm{kt}}$ for $\theta_e \gg 1$ as

$$\mathrm{Re}\chi_e^{\mathrm{kt}} \approx -\frac{\omega_{pe}^2}{\Omega^2}\left[1 + 3\left(\frac{qv_{Te}}{\Omega}\right)^2 + 15\left(\frac{qv_{Te}}{\Omega}\right)^4 + \dots\right] \quad . \quad (78)$$

The dispersion relation of the Langmuir waves is thus given by

$$\varepsilon_p(\Omega) = 1 - \frac{\omega_{pe}^2}{\Omega^2} - 3\frac{q^2\omega_{pe}^2 v_{Te}^2}{\Omega^4} = 0 \quad . \quad (79)$$

Solving Eq. (79) for $\Omega$, we find

$$\Omega_{\mathrm{L}}^2 = \frac{\omega_{pe}^2}{2}\left[1 + \sqrt{1 + \frac{12q^2v_{Te}^2}{\omega_{pe}^2}}\right] \quad . \quad (80)$$

For $qv_{Te} \ll \omega_{pe}$, Eq. (80) reduces to the Bohm – Gross dispersion relation,

$$\Omega_{\mathrm{L}}^2 \approx \omega_{pe}^2 + 3q^2v_{Te}^2 = \omega_{pe}^2(1 + 3q^2\lambda_{De}^2). \quad (81)$$

In the fluid model of plasma dynamics, Eq. (81) can be also derived directly from Eq. (56) with a polytropic closure (57). When the motion of ions is negligible, this equation gives, with $\gamma_e = 3$,

$$\varepsilon_p(\Omega) \approx 1 + \chi_e^{\mathrm{fl}} = 1 - \frac{\omega_{pe}^2}{\Omega^2 - 3q^2v_{Te}^2} = 0 \quad . \quad (82)$$

Solving Eq. (82) for $\Omega$ yields Eq. (81).

The Landau damping of the Langmuir plasma wave is found from the imaginary part of $\chi_e^{\mathrm{kt}}$, $\mathrm{Im}\chi_e^{\mathrm{kt}} = \sqrt{\pi}q^{-2}\lambda_{De}^{-2}\theta_e\exp(-\theta_e^2)$. In the regime of weak damping, with $q\lambda_{De} \ll 1$, the Landau damping is given by $\gamma_L \approx -\sqrt{\pi/8}\,q^{-3}\lambda_{De}^{-3}\omega_{pe}\exp[-q^{-2}\lambda_{De}^{-2}/2 - 3/2]$.

We can now appreciate the physical content and the formal structure of the ideal-fluid nonlinear susceptibility $\chi_p^{(3)}$ in its capacity as a low-order rational approximation to the kinetic-theory solution for the susceptibility derived via a specific closure of the hierarchy of equations for the moments of the pertinent kinetic equation. The leading term in the $\theta_e \ll 1$ expansion of the ideal-fluid solution for the electron-susceptibility constituent of $\chi_p^{(3)}$ [Eq. (68)] recovers the leading, Boltzmann-electron term of its kinetic-theory counterpart [Eq. (67)]. The ideal-fluid solution for the ion-susceptibility component of $\chi_p^{(3)}$, on the other hand, correctly reproduces, in the large-$\theta_i$ approximation, the first and second terms in the large-$\theta_i$ expansion of the real part of the kinetic-theory solution for $\chi_p^{(3)}$ [cf. Eqs. (69) and (70)] and dispersion of plasma waves behind SBS and SRS [Eqs. (72), (76)].

The ideal-fluid model, however, is not intended to describe the imaginary parts of $\chi_\alpha$. Ideal-fluid theories thus fail to correctly account for the damping of laser-driven plasma waves, giving rise to unphysical divergences in nonlinear-response properties near plasma-wave resonances (Figs. 1a – 1f). As a consequence, such theories offer no means to consistently predict stimulated gains $g_s$ and gain bandwidths $\Gamma_s$ in perturbative plasma-enabled stimulated light scattering, necessitating phenomenologically defined damping to compute $g_s$ and $\Gamma_s$. Although the ideal-fluid model of plasma dynamics is insufficient for a self-consistent quantitative analysis of stimulated light scattering in plasmas, it offers valuable physical insights into nonlinear laser – plasma interactions, correctly describing the leading terms the nonlinear susceptibility $\chi_p^{(3)}$ and dispersion relations of plasma waves, including the frequencies of IAW and Langmuir-wave plasma eigenmodes (Figs. 1a – 1f).

## 8. $\chi_p^{(3)}$ physics of stimulated light scattering in plasmas

### 8.1. Landau damping and lineshapes

In Figs. 2a – 2c, we illustrate a typical behavior of the real and imaginary parts of the refractive-index change $\Delta\eta$ induced by a laser driver with a wavelength $\lambda_0$ = 1050 nm and field intensity $I_0$ = $10^{14}$ W/cm$^2$ as functions of $\Omega = \omega_0 - \omega_1$ near the IAW resonance in fully ionized helium with an electron temperature $k_B T_e$ = 250 eV, background electron density $n_{e0} = 0.01 n_c$, where $n_c$ is the critical plasma density, and the ion-to-electron-temperature ratio $T_i/T_e$ varying from 0.1 to 1.0 for the beam-crossing angle $\psi = 20°$. The critical plasma density for $\lambda_0$ = 1050 nm is $n_c \approx$ $10^{21}$ cm$^{-3}$. The unperturbed, zero-field plasma refractive index under these conditions is $\eta_0 \approx 0.995$. The wavenumber of the beat wave for $\psi = 20°$ is $q \approx 2.1$ μm$^{-1}$.

As two most prominent tendencies readily seen in Figs. 2a – 2c, an increase in the ion temperature lowers the peak values of $|\mathrm{Re}\Delta\eta|$ and $|\mathrm{Im}\Delta\eta|$ achieved near IAW resonances and gives rise to significant broadening of these peaks. While at low ion temperatures ($T_i = 0.1T_e$, Fig. 2a), $|\mathrm{Im}\Delta\eta|$ reaches ≈ 8.8·10$^{-3}$ at its IAW peak, for high $T_i$ ($T_i = T_e$, Fig. 2c), the IAW-peak value of $|\mathrm{Im}\Delta\eta|$ is much smaller, $|\mathrm{Im}\Delta\eta| \approx 2.2\cdot10^{-4}$, translating into lower SBS gains [Eqs. (43) – (47)]. As a correlated trend, the full width at half-maximum (FWHM) Γ of the IAW peak of $|\mathrm{Im}\Delta\eta|$ increases from $\approx 0.02\,\Omega_{\mathrm{IAW}}/(qc_s)$ at $T_i = 0.1T_e$ to $\approx 0.56\,\Omega_{\mathrm{IAW}}/(qc_s)$ at $T_i = T_e$.

This behavior of the magnitude and the FWHM of the IAW peak of $|\mathrm{Im}\Delta\eta|$ is fully consistent with the physical picture of Landau damping [Eqs. (64), (77)]. Indeed, as $T_i$ grows from $0.1T_e$ to $T_e$, the thermal velocity of ions increases from $v_{Ti} \approx 98$ km/s at $T_i = 0.1T_e$ to $v_{Ti} \approx 310$ km/s at $T_i = T_e$, while the ideal-fluid ion sound speed grows from $c_s \approx 350$ km/s to ≈ 620 km/s. The ion distribution thus not only broadens with the growth in $T_i$, but also shifts as a whole toward $c_s$, increasing the ion component of the IAW. The higher ion density in the IAW translates into a stronger IAW damping due to a stronger drag force that ions exert on electrons as a part of IAW dynamics. The effect that the increase in the ion temperature has on SBS gain-band broadening is accurately described by the second term in Eq. (77), whose form is manifestly representative of ion Landau damping.

As another physically significant tendency, an increase in the electron temperature decreases field-induced modulation of the electron density, the respective refractive-index change, and the resulting SBS gain. To understand this tendency, we represent Eq. (65) for the $\chi_e$ susceptibility as $\chi_e = q^{-2}\lambda_{De}^{-2}[1 + \theta_e Z(\theta_e)] = 4\pi e^2 n_{e0}(q^2 m_e k_B T_e)^{-1}[1 + \theta_e Z(\theta_e)]$. This $\sim T_e^{-1}$ dependence of $\chi_e$ is imprinted on $\mathcal{K}$ via the self-consistent plasma physics behind $\chi_e$ and $\chi_i$ mixing as expressed by Eqs. (18) and (26), leading to a decrease in the real and imaginary parts of the field-induced refractive index change. This tendency is readily seen via a comparison of Figs. 1 and 2. Specifically, with the laser intensity $I_0$ set at $10^{14}$ W/cm$^2$ and with $\lambda_0$ = 1050 nm, $n_{e0} = 0.01n_c$, $T_i = 0.1T_e$, and $\psi = 20°$, the peak value of $|\mathrm{Im}\Delta\eta|$ reaches ≈ 8.8·10$^{-3}$ for $k_B T_e$ = 250 eV (Fig. 2a) but is limited to ≈ 2.3·10$^{-3}$ for $k_B T_e$ = 1 keV (Fig. 1b).

### 8.2. Effects of collisions

At lower electron and ion temperatures, effects of collisions grow in their significance. To understand this tendency, we resort to standard solutions for the mean free paths $\ell_{\alpha\beta}$ for the collisions of plasma particles of sort $\alpha$ with plasma particles of sort $\beta$, $\ell_{\alpha\alpha} = 3\sqrt{\pi}\, v_{T\alpha}/\nu_{\alpha\alpha}$ and $\ell_{ei} = 3\sqrt{\pi/2}\, v_{Te}/\nu_{ei}$, where $\nu_{ee} = 4\pi n_{e0} e^4 m_e^{-2} v_{Te}^{-3} \ln\Lambda_{ee}$, $\nu_{ei} = 4\pi n_{i0} Z^2 e^4 m_e^{-2} v_{Te}^{-3} \ln\Lambda_{ei}$, and $\nu_{ii} = 4\pi n_{i0} Z^4 e^4 m_i^{-2} v_{Ti}^{-3} \ln\Lambda_{ii}$ are the collision rates and $\ln\Lambda_{\alpha\beta}$ are the regularized Coulomb logarithms for the interactions between plasma particles of sort $\alpha$ and $\beta$.

As the ion temperature $T_i$ increases from $0.1T_e$ to $T_e$, with the electron temperature fixed at $k_B T_e$ = 250 eV in Figs. 2a – 2c, the electron – electron and electron – ion mean free paths, $\ell_{ee} \approx$ 170 μm and $\ell_{ei} \approx$ 60 μm, remain much larger than the inverse beat-wave wavenumber, $q^{-1} \approx$ 0.48 μm, viz., $q\ell_{ee} \approx 340 \gg 1$ and $q\ell_{ei} \approx 120 \gg 1$. Such $q\ell_{ee}$ and $q\ell_{ei}$ products guarantee that electron – electron and electron – ion collisions do not play any significant role in stimulated light scattering by the IAW.

The mean free path for ion – ion collisions, on the other hand, decreases from $\ell_{ii} \approx$ 22 μm to $\ell_{ii} \approx$ 0.44 μm as $T_i$ lowers from $T_e$ to $0.1T_e$, translating into, respectively, $q\ell_{ii} \approx$ 46 at $T_i = T_e$ and $q\ell_{ii} \approx$ 0.91 at $T_i = 0.1T_e$. That $\ell_{ii}$ is on the order of $q^{-1}$ for $T_i = 0.1T_e$ indicates that this regime of laser – plasma interactions, i.e., the regie where $k_B T_e$ = 250 eV, $n_{e0} = 0.01 n_c$, $\lambda_0$ = 1050 nm, $\psi = 20°$, and $T_i = 0.1T_e$, is on the borderline of applicability of collisionless plasma models.

Collisions become less significant at higher electron temperatures. Specifically, with $T_e$ set at 1 keV, i.e., at the level of electron temperatures typical of a vast class of strong-field laser – plasma experiments, including the most recent studies on laser-driven inertial confinement fusion, all the mean free paths $\ell_{\alpha\beta}$ remain much larger than the inverse beat-wave wavenumber $q^{-1}$ within the entire $0.1T_e$-to-$T_e$ range of ion temperatures. Indeed, even with $T_i$ set at $0.1T_e$, all the mean free paths, $\ell_{ee} \approx$ 2.2 mm, $\ell_{ei} \approx$ 0.77 mm, and $\ell_{ii} \approx$ 4.5 μm are much longer than $q^{-1} \approx$ 0.48 μm, viz., $q\ell_{ee} \approx 4.6 \cdot 10^3 \gg 1$ and $q\ell_{ei} \approx 1.6 \cdot 10^3 \gg 1$, and $q\ell_{ii} \approx 9.2 \gg 1$. Effects of collisions are thus negligible within the entire $0.1T_e$-to-$T_e$ range of $T_i$ in this parameter space.

**8.3. Charge separation**

The electron Debye length for $k_B T_e$ = 250 eV and $n_{e0} = 0.01 n_c$ is $\lambda_{De} \approx 0.04$ μm, leading to $q\lambda_{De} \approx 0.08$, indicating that charge-separation effects on the IAW-wavelength scale are negligible, as $\lambda_{De} \ll q^{-1}$, and the plasma is on the $q^{-1}$ scale, to a good approximation, quasineutral. The increase in $q$ and/or $\lambda_{De}$, however, may push laser – IAW physics to the limits of quasineutrality. Specifically, with the electron temperature set at $k_B T_e$ = 1 keV (Figs. 1a – 1f), the $q\lambda_{De}$ product for $\lambda_0$ = 1050 nm and $n_{e0} = 0.01 n_c$ is still small for low beam-crossing angles, viz., $q\lambda_{De} \approx 0.077 \ll 1$ for $\psi = 10°$ and $q\lambda_{De} \approx 0.15 \ll 1$ for $\psi = 20°$, but reaches $q\lambda_{De} \approx 0.88$ for backward SBS, i.e., with $\psi = 180°$.

An increase in $q\lambda_{De}$ shifts the IAW resonance relative to $\Omega_{\text{IAW}} = qc_s$ (Figs. 1a – 1f). This shift is adequately understood in terms of the IAW dispersion as described by Eqs. (72), which establishes $\Omega_{\text{IAW}} = qc_s$ as the $q\lambda_{De} \to 0$ limit of the IAW resonance frequency. The IAW resonance $\Omega_{\text{IAW}}$ is thus controlled by the electron temperature, beam-crossing angle, and initial electron plasma density, suggesting a clear path toward frequency- and gain-tunable plasma SBS photonics.

**8.4. Stimulated gain: plasmas versus highly nonlinear materials**

We see that, with the intensity of the $\lambda_0$ = 1050 nm pump field set at $10^{14}$ W/cm$^2$, corresponding to the normalized laser amplitude $a_0 \approx 9.0\cdot10^{-3}$ and the ponderomotive electron energy averaged over the field cycle $\mathcal{W}_p = m_e c^2 a_0^2/4 \approx 10.3$ eV, the peak value of $|\text{Im}\Delta\eta|$ induced at the Stokes frequency in a plasma with $k_B T_e = 250$ eV and $n_{e0} = 0.01 n_c$ ranges from $\approx 1.2\cdot10^{-3}$ at $T_i = T_e$ to $\approx 8.8\cdot10^{-3}$ at $T_i = 0.1 T_e$ (Fig. 2a). Such $|\text{Im}\Delta\eta|$ values translate into nonlinear susceptibilities with imaginary parts ranging from $\left|\text{Im}\chi_p^{(3)}\right| \approx 2.3\cdot10^{-16}$ to $\approx 1.7\cdot10^{-15}$ cm$^3$/erg and SBS gains $g_s$ from $\approx$ 1.5 to $\approx$ 11 cm/TW.

As the $q\ell_{\alpha\beta}$ analysis in Section 8.2 shows, SBS at $T_i = 0.1 T_e$ in such a laser – plasma interaction setting is demonstrably on the borderline of the collisionless regime. We therefore take $g_s \approx 5$ cm/TW as a representative conservative estimate for the SBS gain in a $k_B T_e = 250$ eV, $n_{e0} = 0.01 n_c$ plasma. As a meaningful benchmark, typical SBS gains attainable with silica fibers, $g_s \approx 3.0$ cm/GW, are $\approx$ 600 times higher than representative IAW-resonant plasma SBS gains (Figs. 1 – 3). Even higher SBS gains are attainable with highly nonlinear liquid-phase and solid-state materials, e.g., $g_s \approx 70$ cm/GW for carbon disulfide ($CS_2$) [100, 102, 147 – 149], and high-pressure gases, $g_s \approx 13.5$ cm/GW for $CO_2$ at ≈ 40 bar [100, 102, 149 – 152].

The range of applications of such high-SBS-gain materials is remarkably broad, spanning from SBS lasers and phase-conjugate mirrors to SBS pulse compressors and beam combiners, but is not readily expandable to strong-field laser science and technologies. The limitations that beam self-focusing and laser-induced ionization impose laser intensities, peak powers, and energy fluences admissible in experiments with such materials are too prohibitive for strong-field applications. Plasma SBS helps alleviate and even remove many of these limitations. Laser intensities and peak powers routinely used in plasma SBS settings are orders of magnitude higher than the laser intensities and peak powers admissible in high-$g_s$ materials. Moreover, with laser intensities typical of plasma SBS/CBET settings, SBS buildup lengths $\mathcal{L}_s = (g_s I_0)^{-1}$ attainable in laser – plasma interactions are significantly shorter than SBS buildup lengths that can be achieved even for highest-$g_s$ materials in their respective range of admissible laser intensities. Specifically, for carbon disulfide, the SBS gain at 1064 nm is $g_s \approx 70$ cm/GW. The Kerr-effect coefficient of $CS_2$ is estimated as $n_2 \approx 3 \cdot 10^{-14}$ cm$^2$/W, translating into the critical power of self-focusing of $P_{cr} \approx 35$ kW for 1064-nm radiation. The laser pulse width in SBS experiments in $CS_2$ is limited from below by the acoustic buildup time, $\tau_a \approx 6.5$ ns. For a typical pulse width of ≈ 30 ns [153 – 155], self-focusing sets a limit for the pulse energy at ≈ 1.1 mJ. Furthermore, to avoid excessive photoionization, the laser intensity in $CS_2$ needs to be kept below 100 MW/cm$^2$. These factors limit the SBS buildup length is thus limited at $\mathcal{L}_s \approx 0.14$ cm.

Plasma SBS operates with laser pulses of drastically higher pulse energies (hundreds of joules [88 – 97]) and radically higher field intensities ($10^{13}$ to $10^{15}$ W/cm$^2$). With $I_0$ set at $10^{14}$ W/cm$^2$ as a representative midrange estimate, plasma SBS gain $g_s \approx 5$ cm/TW, attainable in the above-specified laser – plasma interaction setting, leads to SBS buildup within a typical length of $\mathcal{L}_s \approx 20$ μm. Thus, even though plasma SBS gains are orders of magnitude lower than the SBS gains attainable with highly nonlinear materials, SBS in suitably tailored plasmas not only admits much higher pulse energies and field intensities, but also builds up on much shorter length scales compared to SBS in high-$g_s$ materials.

## 9. Conclusion

To summarize, we have shown that, within a vast parameter space of laser – plasma physics, including practically significant regimes of strong-field laser – plasma interactions, the notion of nonlinear-optical susceptibility permits a heuristically valuable extension to stimulated light scattering in plasmas, offering powerful insights into the properties of stimulated Brillouin and stimulated Raman scattering from plasma waves and shedding new light on the inner workings of the related cross-beam energy transfer processes. We have derived a closed-form, physically intuitive solution for the third-order susceptibility $\chi_p^{(3)}$ for stimulated Brillouin and stimulated Raman scattering, expressing $\chi_p^{(3)}$ in terms of individual susceptibilities $\chi_e$ and $\chi_i$ of electron and ion plasma constituents. While the individual susceptibilities $\chi_e$ and $\chi_i$ are functions of plasma parameters, depending on the specific regime of laser – plasma interactions, the algebra whereby $\chi_e$ and $\chi_i$ are mixed into $\chi_p^{(3)}$ remains unchanged in a vast variety of physical settings, revealing significant physical properties of plasma SBS and SRS, including a physically substantiative connection between SBS/SRS spectral lineshapes and Landau damping, and allowing the ideal-fluid cubic susceptibility $\chi_p^{(3)}$ to be understood as a low-order rational approximation to the kinetic-theory solution for $\chi_p^{(3)}$ derived via a specific closure of the hierarchy of equations for the moments of the governing kinetic equation. We have shown that, in its weak-field, steady-state limit, the general physical picture of plasma SBS and SRS that views these processes as Stokes-field – plasma-wave decay instabilities of the pump field, reduces to a perturbative picture, familiar from nonlinear optics of neutral media, in which constitutive relations for plasma SBS/SRS response are formulated in terms of the third-order nonlinear polarization and the SBS/SRS gain is defined by the imaginary part of $\chi_p^{(3)}$. These findings lay grounds for a unified treatment of stimulated light scattering across a vast area of optical physics, spanning from fiber optics and neutral gases to laser – plasma physics.

**Acknowledgments**

This research was supported in part by the DOE Office of Science through the DOE Inertial Fusion Energy Science and Technology Accelerator Research (IFE-STAR) program (grant # DE-SC0024882).

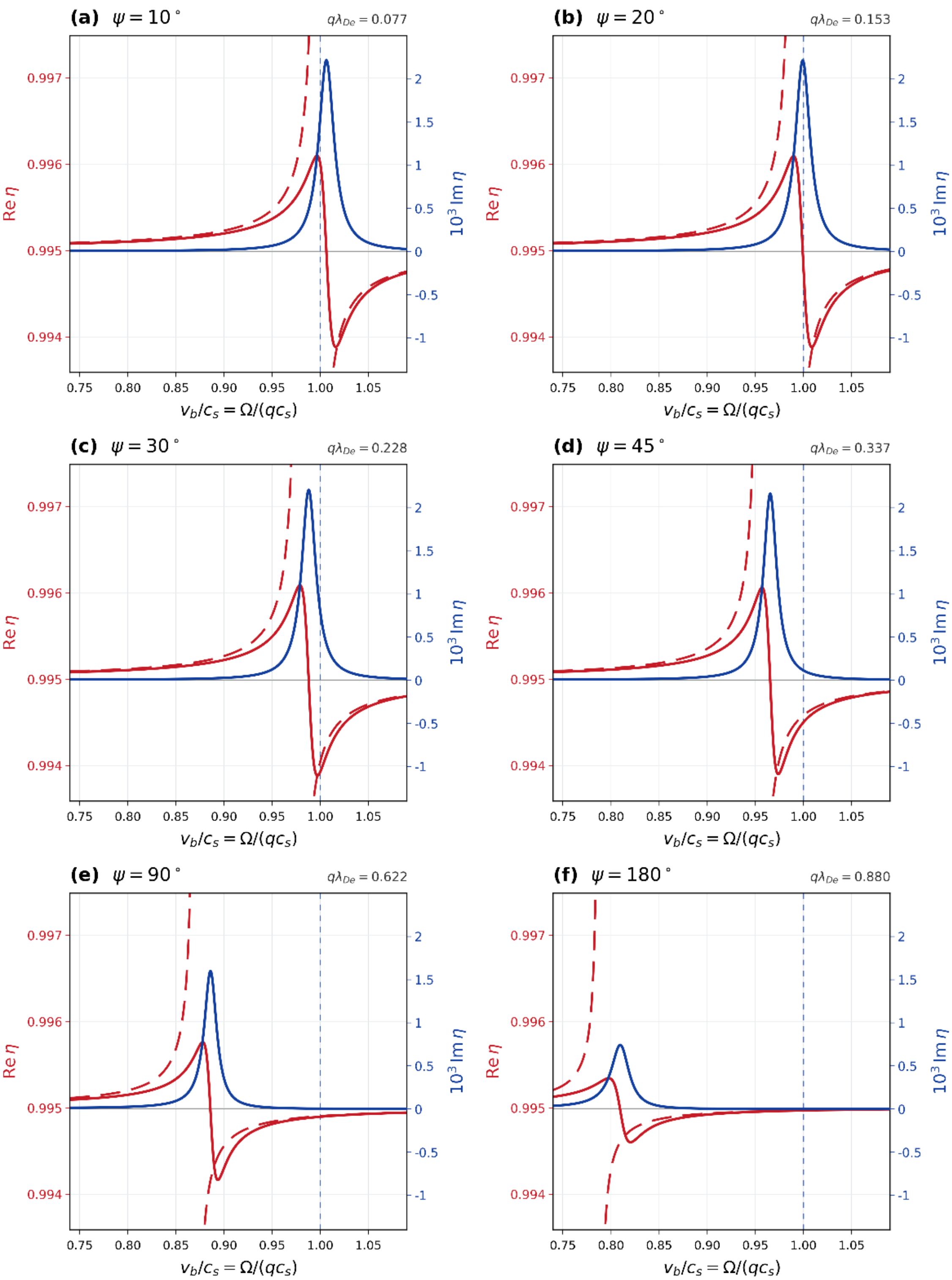

Fig. 1. The real (red curves) and imaginary (blue curves) parts of the field-induced refractive index change as functions of $v_b/c_s = \Omega/(qc_s)$ calculated using the kinetic (solid lines) and ideal-fluid (dashed lines) models of laser – plasma interactions for a helium plasma with an electron temperature $k_B T_e$ = 1 keV, ion temperature $k_B T_i$ = 100 eV, and background electron density $n_{e0} = 0.01 n_c$ driven by a laser field with a wavelength $\lambda_0$ = 1050 nm, field intensity $I_0$ = $10^{14}$ W/cm$^2$, and pump – Stokes beam-crossing angle $\psi = 10°$ (a), $20°$ (b), $30°$ (c), $45°$ (d), $90°$ (e), and $180°$ (f).

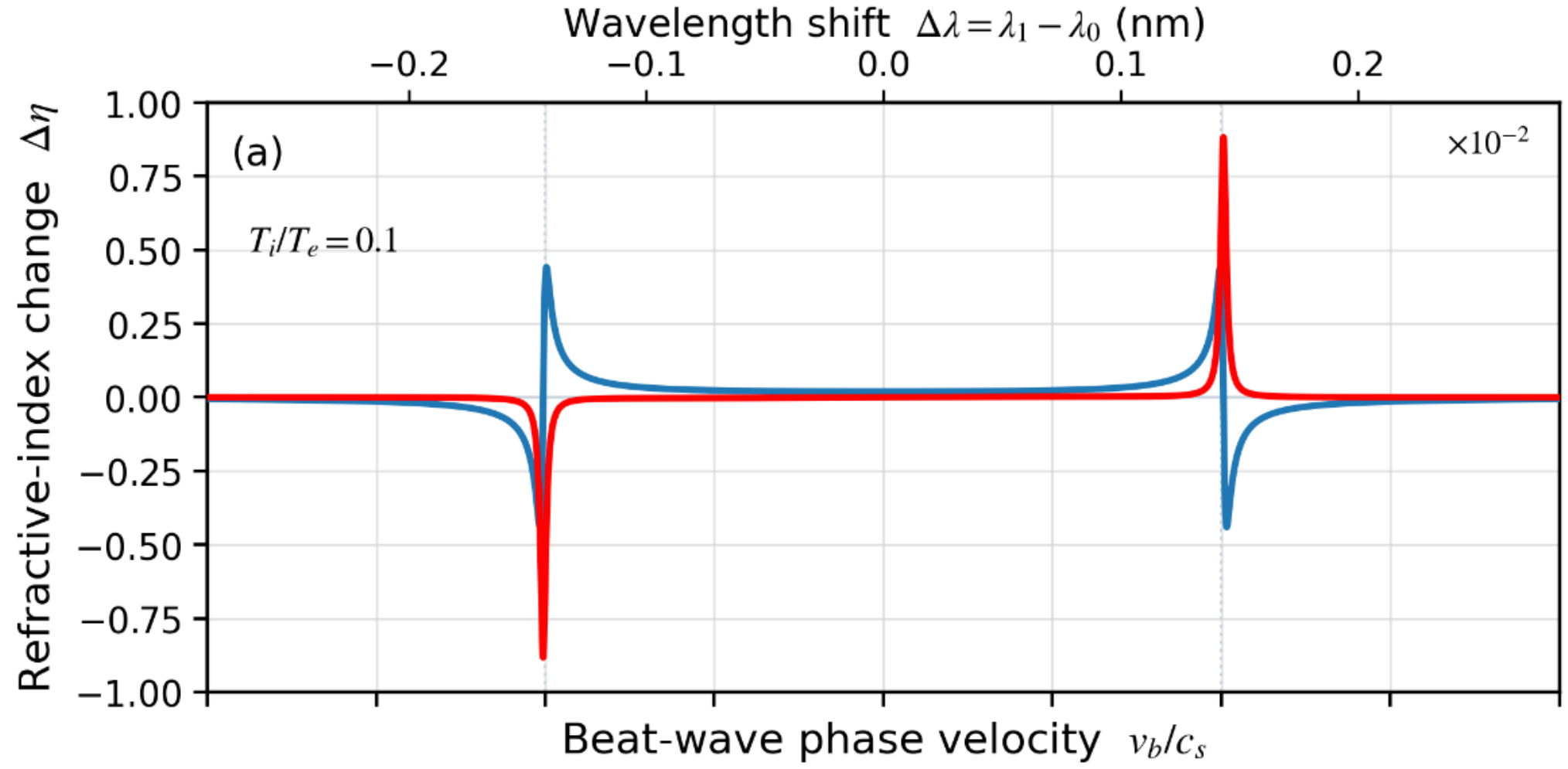

Wavelength shift $\Delta\lambda = \lambda_1 - \lambda_0$ (nm)
−0.2
−0.1
0.0
0.1
0.2
1.00
0.75
0.50
0.25
0.00
−0.25
−0.50
−0.75
−1.00
(a)
$\times 10^{-2}$
$T_i/T_e = 0.1$
Refractive-index change $\Delta\eta$
Beat-wave phase velocity $v_b/c_s$

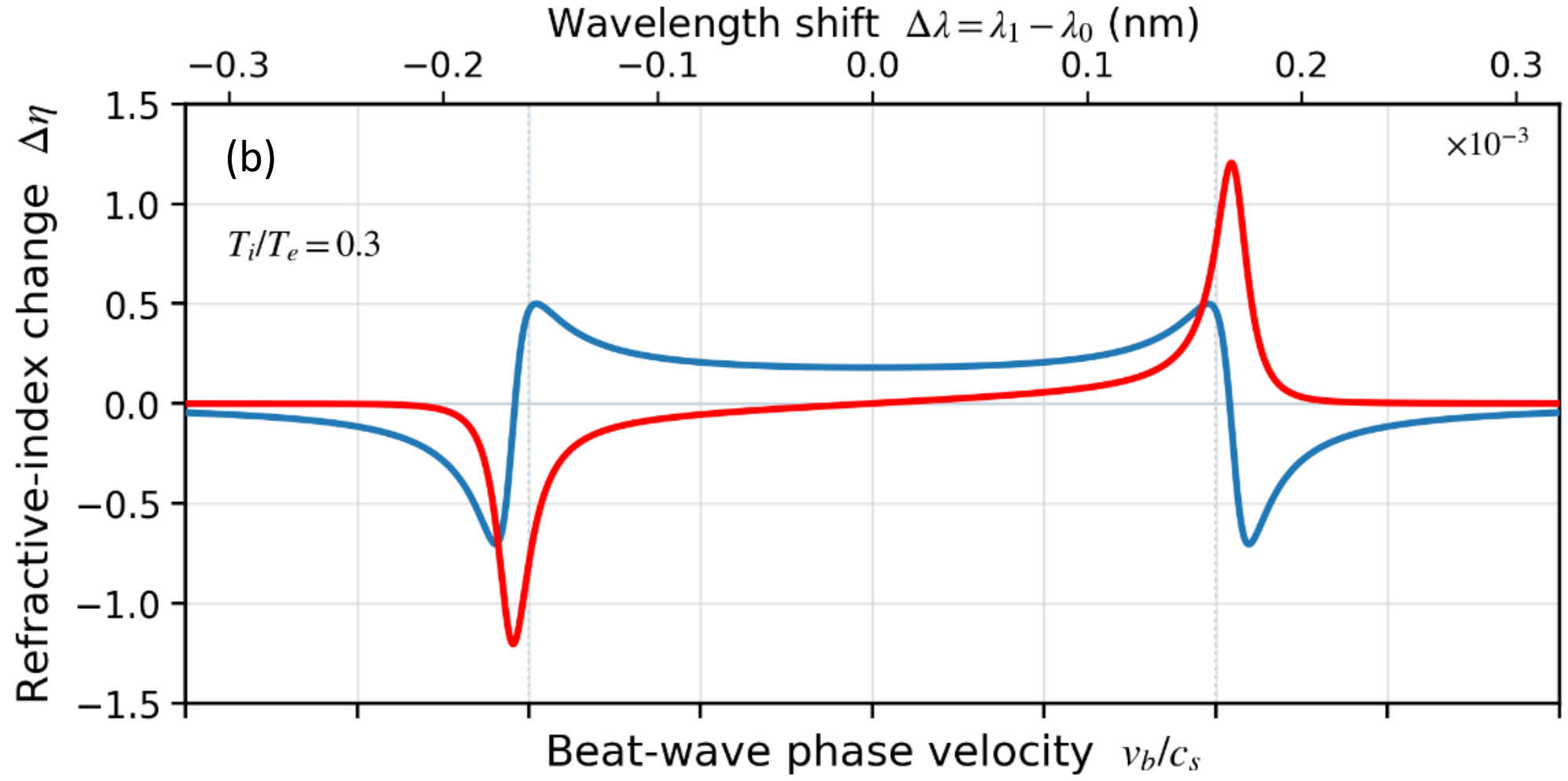

Wavelength shift $\Delta\lambda = \lambda_1 - \lambda_0$ (nm)
−0.3
−0.2
−0.1
0.0
0.1
0.2
0.3
1.5
1.0
0.5
0.0
−0.5
−1.0
−1.5
(b)
$\times 10^{-3}$
$T_i/T_e = 0.3$
Refractive-index change $\Delta\eta$
Beat-wave phase velocity $v_b/c_s$

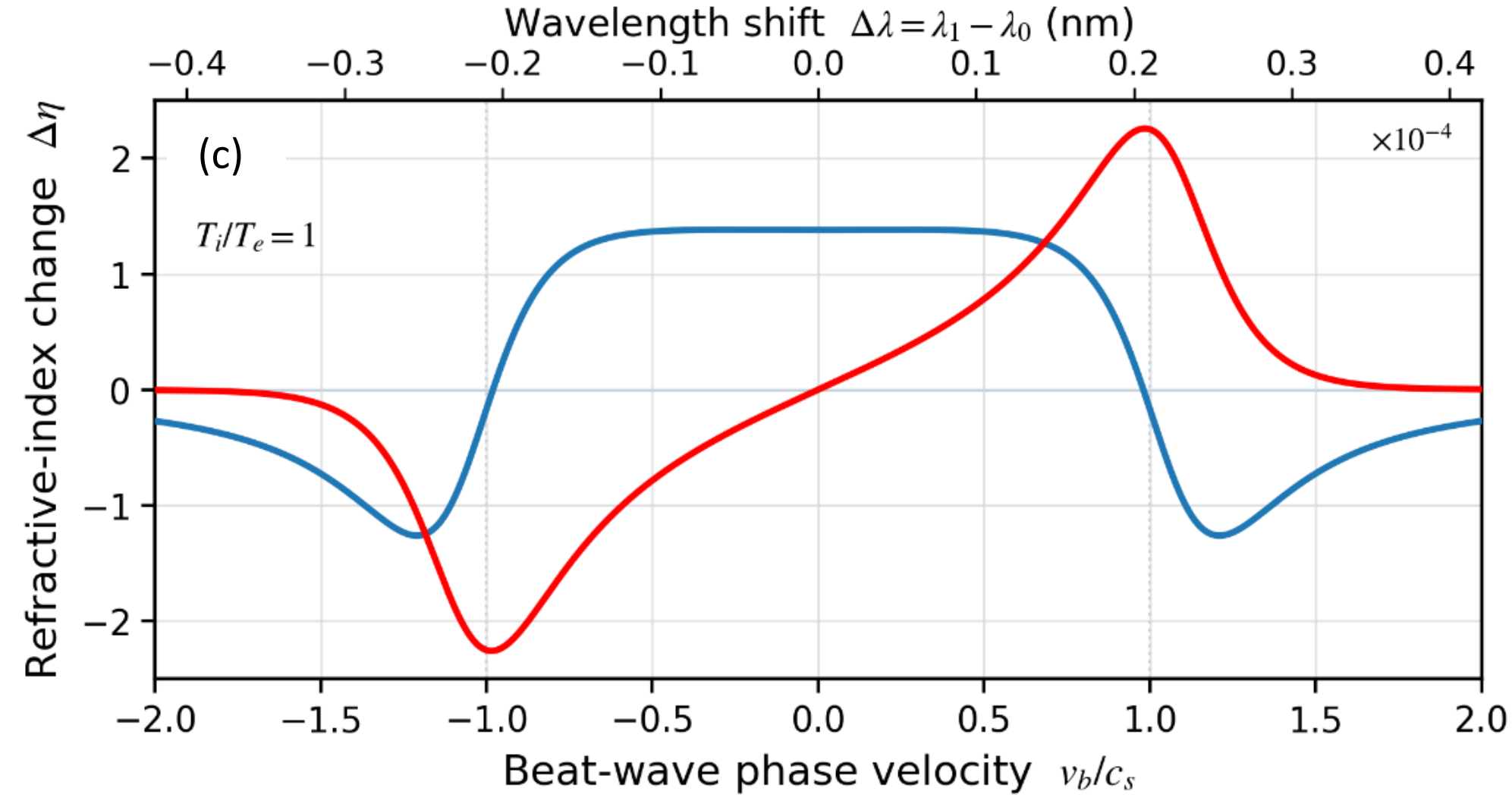

Wavelength shift $\Delta\lambda = \lambda_1 - \lambda_0$ (nm)
−0.4
−0.3
−0.2
−0.1
0.0
0.1
0.2
0.3
0.4
2
1
0
−1
−2
(c)
$\times 10^{-4}$
$T_i/T_e = 1$
Refractive-index change $\Delta\eta$
−2.0
−1.5
−1.0
−0.5
0.0
0.5
1.0
1.5
2.0
Beat-wave phase velocity $v_b/c_s$

Fig. 2. The real (blue curves) and imaginary (red curves) parts of the field-induced refractive index change as functions of $v_b/c_s = \Omega/(qc_s)$ for a helium plasma with an electron temperature $k_BT_e$ = 250 eV and background electron density $n_{e0} = 0.01n_c$ driven by a laser field with a wavelength $\lambda_0$ = 1050 nm, field intensity $I_0 = 10^{14}$ W/cm$^2$, and pump – Stokes beam-crossing angle $\psi = 20°$. The ratio of the ion temperature to the electron temperature is $T_i/T_e$ = 0.1 (a), 0.3 (b), and 1.0 (c).

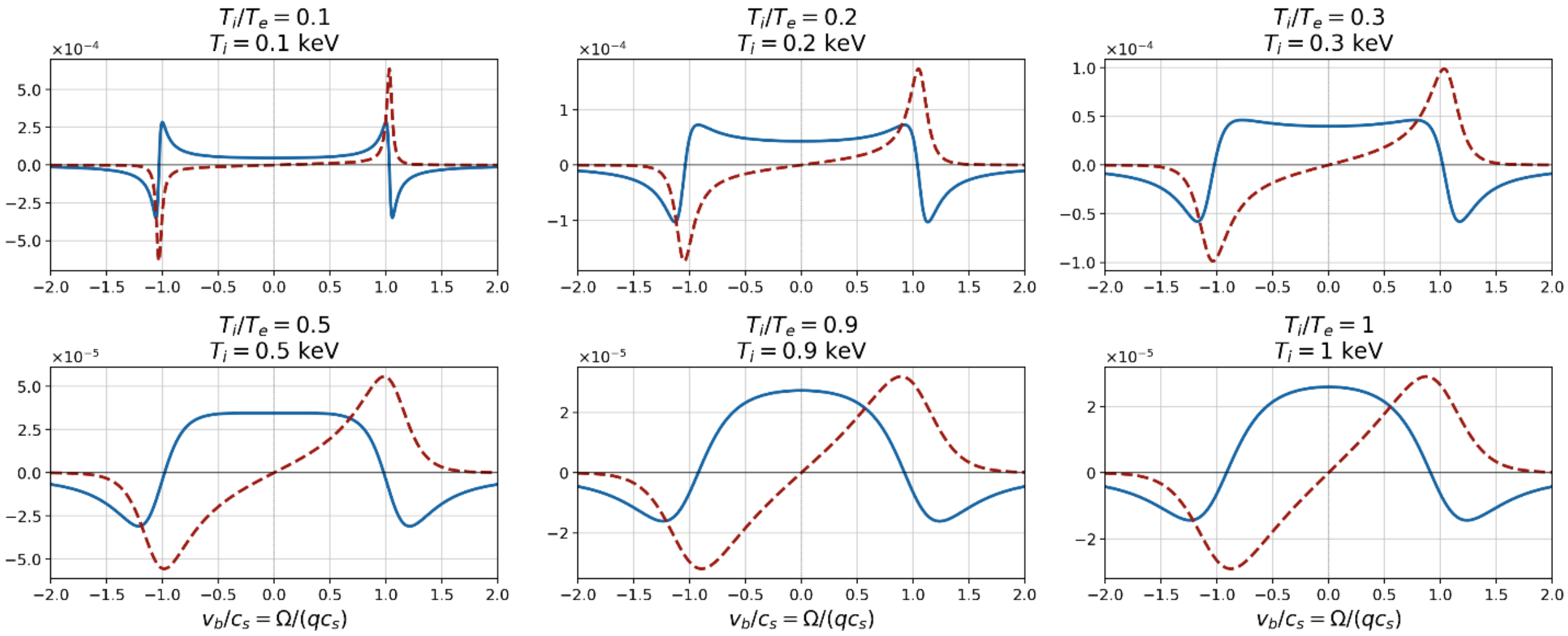

Fig. 3. The real (blue solid curves) and imaginary (red dashed curves) parts of the field-induced refractive index change as functions of $v_b/c_s = \Omega/(qc_s)$ for a hydrogen plasma with an electron temperature $k_BT_e = 1$ keV and background electron density $n_{e0} = 0.01n_c$ driven by a laser field with a wavelength $\lambda_0 = 1050$ nm, field intensity $I_0 = 10^{14}$ W/cm$^2$, and pump – Stokes beam-crossing angle $\psi = 10°$. The ratio of the ion temperature to the electron temperature is $T_i/T_e =$ 0.1 (a), 0.2 (b), 0.3 (c), 0.5 (d), 0.9 (e), and 1.0 (f).